\documentclass[12pt]{article}
\usepackage{lineno}
\usepackage{graphicx}
\usepackage{amsmath}
\usepackage{hhline}
\usepackage{amssymb}
\usepackage{mathptmx}
\usepackage{cite}
\usepackage{array}
\usepackage{multirow}
\usepackage{siunitx}
\usepackage{textcomp}
\usepackage{rotating}
\usepackage{tabularx}
\usepackage{mathtools}
\usepackage{placeins}

\usepackage{color}
\definecolor{rltred}{rgb}{0.75,0,0}
\definecolor{rltgreen}{rgb}{0,0.5,0}
\definecolor{rltblue}{rgb}{0,0,0.75}

\newif\ifpdf
\ifx\pdfoutput\undefined
    \pdffalse          
\else
    \pdfoutput=1       
    \pdftrue
\fi

\ifpdf
\usepackage{thumbpdf}
\usepackage[pdftex,
        colorlinks=true,
        urlcolor=rltblue,       
        filecolor=rltgreen,     
        linkcolor=rltred,       
        pdftitle={Example File for pdflatex},
        pdfauthor={H1},
        pdfsubject={Template file},
        pdfkeywords={High-Energy Physics, Particle Physics},
        pdfpagemode=None,
        bookmarksopen=true]{hyperref}
 
\fi

\newlength{\dinwidth}
\newlength{\dinmargin}
\begin{document}  

\newcommand{\be}{\begin{equation}}
\newcommand{\ee}{\end{equation}}
\newcommand{\bea}{\begin{eqnarray}}
\newcommand{\eea}{\end{eqnarray}}

\newcommand{\pom}{{I\!\!P}}
\newcommand{\reg}{{I\!\!R}}
\newcommand{\slowpi}{\pi_{\mathit{slow}}}
\newcommand{\fiidiii}{F_2^{D(3)}}
\newcommand{\fiidiiiarg}{\fiidiii\,(\beta,\,Q^2,\,x)}
\newcommand{\n}{1.19\pm 0.06 (stat.) \pm0.07 (syst.)}
\newcommand{\nz}{1.30\pm 0.08 (stat.)^{+0.08}_{-0.14} (syst.)}
\newcommand{\fiidiiiful}{F_2^{D(4)}\,(\beta,\,Q^2,\,x,\,t)}
\newcommand{\fiipom}{\tilde F_2^D}
\newcommand{\ALPHA}{1.10\pm0.03 (stat.) \pm0.04 (syst.)}
\newcommand{\ALPHAZ}{1.15\pm0.04 (stat.)^{+0.04}_{-0.07} (syst.)}
\newcommand{\fiipomarg}{\fiipom\,(\beta,\,Q^2)}
\newcommand{\pomflux}{f_{\pom / p}}
\newcommand{\nxpom}{1.19\pm 0.06 (stat.) \pm0.07 (syst.)}
\newcommand {\gapprox}
   {\raisebox{-0.7ex}{$\stackrel {\textstyle>}{\sim}$}}
\newcommand {\lapprox}
   {\raisebox{-0.7ex}{$\stackrel {\textstyle<}{\sim}$}}
\def\gsim{\,\lower.25ex\hbox{$\scriptstyle\sim$}\kern-1.30ex%
\raise 0.55ex\hbox{$\scriptstyle >$}\,}
\def\lsim{\,\lower.25ex\hbox{$\scriptstyle\sim$}\kern-1.30ex%
\raise 0.55ex\hbox{$\scriptstyle <$}\,}
\newcommand{\pomfluxarg}{f_{\pom / p}\,(x_\pom)}
\newcommand{\dsf}{\mbox{$F_2^{D(3)}$}}
\newcommand{\dsfva}{\mbox{$F_2^{D(3)}(\beta,Q^2,x_{I\!\!P})$}}
\newcommand{\dsfvb}{\mbox{$F_2^{D(3)}(\beta,Q^2,x)$}}
\newcommand{\dsfpom}{$F_2^{I\!\!P}$}
\newcommand{\gap}{\stackrel{>}{\sim}}
\newcommand{\lap}{\stackrel{<}{\sim}}
\newcommand{\fem}{$F_2^{em}$}
\newcommand{\tsnmp}{$\tilde{\sigma}_{NC}(e^{\mp})$}
\newcommand{\tsnm}{$\tilde{\sigma}_{NC}(e^-)$}
\newcommand{\tsnp}{$\tilde{\sigma}_{NC}(e^+)$}
\newcommand{\st}{$\star$}
\newcommand{\sst}{$\star \star$}
\newcommand{\ssst}{$\star \star \star$}
\newcommand{\sssst}{$\star \star \star \star$}
\newcommand{\tw}{\theta_W}
\newcommand{\sw}{\sin{\theta_W}}
\newcommand{\cw}{\cos{\theta_W}}
\newcommand{\sww}{\sin^2{\theta_W}}
\newcommand{\cww}{\cos^2{\theta_W}}
\newcommand{\trm}{m_{\perp}}
\newcommand{\trp}{p_{\perp}}
\newcommand{\trmm}{m_{\perp}^2}
\newcommand{\trpp}{p_{\perp}^2}
\newcommand{\alp}{\alpha_s}

\newcommand{\alps}{\alpha_s}
\newcommand{\sqrts}{$\sqrt{s}$}
\newcommand{\LO}{$O(\alpha_s^0)$}
\newcommand{\Oa}{$O(\alpha_s)$}
\newcommand{\Oaa}{$O(\alpha_s^2)$}
\newcommand{\PT}{p_{\perp}}
\newcommand{\JPSI}{J/\psi}
\newcommand{\sh}{\hat{s}}
\newcommand{\uh}{\hat{u}}
\newcommand{\MP}{m_{J/\psi}}
\newcommand{\PO}{I\!\!P}
\newcommand{\xbj}{$x_{\rm bj}$}
\newcommand{\xpom}{x_{\PO}}
\newcommand{\ttbs}{\char'134}
\newcommand{\xpomlo}{3\times10^{-4}}  
\newcommand{\xpomup}{0.05}  
\newcommand{\dgr}{^\circ}
\newcommand{\pbarnt}{\,\mbox{{\rm pb$^{-1}$}}}
\newcommand{\gev}{\,\mbox{GeV}}
\newcommand{\mev}{\,\mbox{MeV}}
\newcommand{\WBoson}{\mbox{$W$}}
\newcommand{\fbarn}{\,\mbox{{\rm fb}}}
\newcommand{\fbarnt}{\,\mbox{{\rm fb$^{-1}$}}}

%
%
\newcommand{\qsq}{\ensuremath{Q^2} }
\newcommand{\qsqnospace}{\ensuremath{Q^2}}
\newcommand{\gevsq}{\ensuremath{\mathrm{GeV}^2} }
\newcommand{\mevc}{\ensuremath{\mathrm{MeV/c}} }
\newcommand{\et}{\ensuremath{E_t^\ast} }
\newcommand{\rap}{\ensuremath{\eta^\ast} }
\newcommand{\gp}{\ensuremath{\gamma^\ast}p }
\newcommand{\dsiget}{\ensuremath{{\rm d}\sigma_{ep}/{\rm d}E_t^\ast} }
\newcommand{\dsigrap}{\ensuremath{{\rm d}\sigma_{ep}/{\rm d}\eta^\ast} }
\def\Journal#1#2#3#4{{#1} {\bf #2} (#3) #4}
\def\NCA{\em Nuovo Cimento}
\def\NIM{\em Nucl. Instrum. Methods}
\def\NIMA{{\em Nucl. Instrum. Methods} {\bf A}}
\def\NPB{{\em Nucl. Phys.}   {\bf B}}
\def\PLB{{\em Phys. Lett.}   {\bf B}}
\def\PRL{\em Phys. Rev. Lett.}
\def\PRD{{\em Phys. Rev.}    {\bf D}}
\def\ZPC{{\em Z. Phys.}      {\bf C}}
\def\EJC{{\em Eur. Phys. J.} {\bf C}}
\def\CPC{\em Comp. Phys. Commun.}

\begin{titlepage}

\noindent
\begin{flushleft}
{\tt DESY 20-176    \hfill    ISSN 0418-9833} \\
{\tt Oct 2020}                  \\
\end{flushleft}

\noindent
\vspace{2cm}

\begin{center}
\begin{Large}

{\bf Measurement of charged particle multiplicity distributions in DIS at HERA and its implication to entanglement entropy of partons}

\vspace{2cm}

H1 Collaboration

\end{Large}
\end{center}

\vspace{2cm}

\begin{abstract}
Charged particle multiplicity distributions in positron-proton deep inelastic scattering at a centre-of-mass energy $\sqrt{s}=319$\gev\ are
measured. The data are collected with the H1 detector at HERA corresponding to an integrated luminosity of $136$~\pbarnt.
Charged particle multiplicities are measured as a function of photon virtuality \qsqnospace, inelasticity $y$ and pseudorapidity $\eta$ in the laboratory and the hadronic centre-of-mass frames. Predictions from different Monte Carlo models are compared to the data. The first and second moments of the multiplicity distributions are determined and the KNO scaling behaviour is investigated. The multiplicity distributions as a function of \qsq and the Bjorken variable \xbj{} are converted to the hadron entropy $S_{\rm hadron}$, and predictions from a quantum entanglement model are tested.

\end{abstract}

\vspace{1.5cm}

\begin{center}
To be submitted to EPJC
\end{center}

\end{titlepage}

%
%
%
\begin{flushleft}

V.~Andreev$^{19}$,             
A.~Baghdasaryan$^{30}$,        
A.~Baty$^{45}$,                
K.~Begzsuren$^{27}$,           
A.~Belousov$^{19}$,            
A.~Bolz$^{12}$,                
V.~Boudry$^{22}$,              
G.~Brandt$^{40}$,              
D.~Britzger$^{20}$,            
A.~Buniatyan$^{2}$
% ...


\bigskip{\it
% ...
$ ^{2}$ School of Physics and Astronomy, University of Birmingham,
          Birmingham, UK$^{ b}$ \\
% ...
$ ^{12}$ Physikalisches Institut, Universit\"at Heidelberg,
          Heidelberg, Germany$^{ a}$ \\
% ...
$ ^{19}$ Lebedev Physical Institute, Moscow, Russia \\
 $ ^{20}$ Max-Planck-Institut f\"ur Physik, M\"unchen, Germany \\
% ...
$ ^{22}$ LLR, Ecole Polytechnique, CNRS/IN2P3, Palaiseau, France \\
% ...
$ ^{27}$ Institute of Physics and Technology of the Mongolian
          Academy of Sciences, Ulaanbaatar, Mongolia \\
% ...
$ ^{30}$ Yerevan Physics Institute, Yerevan, Armenia \\
% ...
$ ^{40}$ II. Physikalisches Institut, Universit\"at G\"ottingen,
          G\"ottingen, Germany \\
% ...
$ ^{45}$ Rice University, Houston, USA \\
% ...
$ ^a$ Supported by the Bundesministerium f\"ur Bildung und Forschung, FRG,
      under contract numbers 05H09GUF, 05H09VHC, 05H09VHF,  05H16PEA \\
 $ ^b$ Supported by the UK Science and Technology Facilities Council,
      and formerly by the UK Particle Physics and
      Astronomy Research Council \\
% ...
}
\end{flushleft}
\newpage

\section{\label{sec:intro}Introduction}
\noindent
In the parton model\cite{Bjorken:1969ja,Feynman:1969wa,Gribov:1968fc} formulated by Bjorken, Feynman, and Gribov, the  quarks and gluons of a nucleon are viewed as ``quasi-free" particles probed by an external hard probe in the infinite momentum frame. The parton that participates in the hard interaction with the probe, e.g., a virtual photon, is expected to be causally disconnected from the rest of the nucleon. On the other hand, the parton and the rest of the nucleon have to form a colour-singlet state due to colour confinement. In order to further understand the role of colour confinement in high energy collisions, it has been suggested~\cite{Klebanov:2007ws,Kharzeev:2017qzs} that quantum entanglement of partons could be an important probe of the underlying mechanism. 
Measurements of charged particle multiplicities are proposed~\cite{Kharzeev:2017qzs,Duan:2020jkz,Armesto:2019mna,Kovner:2015hga,Baker:2017wtt,Tu:2019ouv} to be related to the entanglement entropy predicted from the gluon density~\cite{Kharzeev:2017qzs}, as an indication of quantum entanglement of partons inside the proton. 

Particle multiplicity distributions have previously been measured in deep inelastic scattering (DIS) at HERA~\cite{Derrick:1995ca,Aid:1996cb,Adloff:1997fr,Adloff:1998dw,Chekanov:2008ae}. 
In this paper, charged particle multiplicity distributions in positron-proton ($ep$) DIS at $\sqrt{s}=319$\gev~are reported using high statistics data collected with the H1 detector.
The phase space of the measurement of multiplicity distributions is defined in bins of the virtuality of the photon $5<\qsq<100 ~\gevsq$, the inelasticity $y$ variable $0.0375<y<0.6$ and the pseudorapidity $\eta$ of charged particles, $|\eta_{_{\text{lab}}}|<1.6$ in the laboratory frame, and $0<\eta^{\ast}<4$ in the hadronic centre-of-mass (HCM) frame.\footnote{The positive $\eta^{\ast}$ region qualitatively corresponds to the current fragmentation region (predominantly negative $\eta_{_{\rm lab}}$), while $\eta^{\ast}<0$ corresponds to the direction of the target fragmentation region (predominantly positive $\eta_{_{\rm lab}}$).} The first and the second moments of the multiplicity distributions, corresponding to the mean and the variance, are measured as a function of the hadronic centre-of-mass energy $W$, in different bins of \qsqnospace.
The KNO scaling function~\cite{Koba:1972ng} is also measured in different $W$ and \qsq regions.  

The final-state hadron entropy, $S_{\rm hadron}$, as a function of the Bjorken variable \xbj~in different \qsq bins is also measured. A relation between the $S_{\rm hadron}$ and the initial-state parton entropy, $S_{\rm gluon}$, due to ``parton liberation"~\cite{Mueller:1999fp} and ``local parton-hadron duality (LPHD)"~\cite{Dokshitzer_1991}, is described by~\cite{Kharzeev:2017qzs}:
\begin{linenomath*}
\be\label{eq1}
S_{\rm hadron} \equiv -\sum{P(N)\ln{P(N)}}=\ln{[xG(x,\qsq)]} \equiv S_{\rm gluon}.
\ee
\end{linenomath*}
\noindent Here, $P(N)$ is the charged particle multiplicity distribution measured in either the current fragmentation region or the target fragmentation region, where the gluon density $xG(x,Q^{2})$ is evaluated at $x =$~\xbj.

The measurements are compared to theoretical predictions obtained from simulations including parton showers and hadronisation (RAPGAP~\cite{Jung:1993gf}, DJANGOH~\cite{Charchula:1994kf} and PYTHIA 8~\cite{Sjostrand:2014zea}).

\section{\label{subsec:h1} H1 detector}

A full description of the H1 detector can be found elsewhere~\cite{Abt:1996hi,Abt:1996xv,Appuhn:1996na,Pitzl:2000wz,List:2002bd,Becker:2007ms,Laycock:2012xg} and only the components
most relevant for this analysis are briefly mentioned here. The coordinate system of H1 is
defined such that the positive $z$ axis is pointing in the proton beam direction (forward
direction) and the nominal interaction point is located at $z$ = 0. The polar angle $\theta$ is 
defined with respect to this axis. The pseudorapidity is defined to be $\eta \equiv -\ln{(\tan{(\theta/2)})}$.

Charged particles are measured in the polar angle range $\ang{15}<\theta<\ang{165}$ using the central tracking detector (CTD), which is also used to reconstruct the interaction vertex. The CTD
comprises two large cylindrical, concentric and coaxial jet chambers (CJCs), and the silicon vertex detector~\cite{Pitzl:2000wz,List:2002bd}. The CTD is operated inside a 1.16 T solenoidal magnetic field. The CJCs are separated by a cylindrical drift chamber which improves the $z$ coordinate reconstruction. A cylindrical multiwire proportional chamber~\cite{Becker:2007ms}, which is mainly used in the trigger, is situated inside the inner CJC. The trajectories of charged particles are measured with a transverse momentum resolution
of $\sigma(p_{T})/p_{T} \approx 0.2\%/\gev \oplus 1.5\%$. The forward tracking detector (FTD)~\cite{Laycock:2012xg} measures the tracks of charged particles at polar angles $\ang{6}<\theta<\ang{25}$. In the region of angular overlap, FTD and short CTD track segments are used to reconstruct combined tracks, extending the detector acceptance for well-reconstructed tracks. Both CTD tracks and combined tracks are linked to hits in the vertex detectors: the central silicon tracker (CST)~\cite{Pitzl:2000wz,List:2002bd}, the backward silicon tracker (BST) and the forward silicon tracker (FST). These detectors provide precise spatial coordinate measurements and therefore significantly improve the primary vertex spatial resolution. The CST consists of two layers of double-sided silicon strip detectors surrounding the beam pipe covering an angular range of $\ang{30}<\theta<\ang{150}$ for tracks passing through both layers. The BST consists of six double wheels of silicon strip detectors measuring the transverse coordinates of charged particles. The FST design is similar to that of the BST and consists of five double wheels of single-sided silicon strip detectors. The lead-scintillating fibre calorimeter (SpaCal)~\cite{Appuhn:1996na} covering the region $\ang{153}<\theta<\ang{177.5}$ has electromagnetic and hadronic sections. The calorimeter is used to measure the scattered positron and the backward hadronic energy flow. The energy resolution for positrons in the electromagnetic section is $\sigma(E)/E \approx 7.1\%/ \sqrt{E/\gev} \oplus 1\%$, as determined in test beam measurements~\cite{Nicholls:1995di}. The SpaCal provides energy and time-of-flight information used for triggering purposes. A backward proportional chamber (BPC) in front of the SpaCal is used to improve the angular measurement of the scattered lepton. The liquid argon (LAr) calorimeter~\cite{Andrieu:1993kh} covers the range $\ang{4}<\theta<\ang{154}$ and is used in this analysis in the reconstruction of the hadronic final state. It has an energy resolution of $\sigma(E)/E \approx 50\%/ \sqrt{E/\gev} \oplus 2\%$ for hadronic showers, as obtained from test beam measurements~\cite{Andrieu:1993tz}.

\section{\label{sec:model} Theoretical predictions}

The DIS process is simulated by different Monte Carlo (MC) event generators, which include the hard scattering process and simulations of higher order QCD correction in form of parton showers and hadronisation. A brief description of the MC event generators is given below: 

\begin{itemize}

\item The \textbf{RAPGAP 3.1}~\cite{Jung:1993gf} MC event generator matches first order Quantum Chromodynamics (QCD) matrix elements to the Dokshitzer-Gribov-Lipatov-Altarelli-Parisi (DGLAP)\\\cite{Gribov:1972ri,Lipatov:1974qm,Altarelli:1977zs,Dokshitzer:1977sg} parton showers with strongly ordered transverse momenta of subsequently emitted partons. The factorisation and renormalisation scales are set to $\mu_{f}=\mu_{r}=\sqrt{\qsq+\hat{p}^{2}_{T}}$, where $\hat{p}_{T}$ is the transverse momentum of the outgoing hard parton from the matrix element in the centre-of-mass frame of the hard subsystem. The CTEQ 6L~\cite{Pumplin:2002vw} leading order parametrisation of the parton density function (PDF) is used.

\item The \textbf{DJANGOH 1.4}~\cite{Charchula:1994kf} MC event generator uses the Colour Dipole Model (CDM) as implemented in ARIADNE~\cite{Lonnblad:1992tz}, which models first order QCD processes and creates dipoles between coloured partons. Gluon emission is treated as radiation from these dipoles, and new dipoles are formed from the emitted gluons from which further radiation is possible. The radiation pattern of the dipoles includes interference effects, thus modelling gluon coherence. The transverse momenta of the emitted partons are not ordered in transverse momentum with respect to rapidity, producing a configuration similar to the  Balitsky-Fadin-Kuraev-Lipatov (BFKL)~\cite{Kuraev:1976ge,Kuraev:1977fs,Balitsky:1978ic} treatment of parton evolution~\cite{Lonnblad:1994wk}. The CTEQ 6L~\cite{Pumplin:2002vw} at leading order is used as the PDF. 

\item The \textbf{PYTHIA 8}~\cite{Sjostrand:2014zea} MC event generator models the hard collision by LO-pQCD cross sections. In order to enable the PYTHIA 8 program to simulate DIS processes, the ${\cal O}(\alpha_s^0)$ process together with the DIrect REsummation (DIRE) parton shower~\cite{Hoche:2015sya} is applied. The parton shower recoil is treated such that the kinematic variables of the DIS process ($y$ and $\qsq$) are unchanged. The PDF CTEQ 5L~\cite{Pumplin:2002vw} is used. 
\end{itemize}

DJANGOH and RAPGAP, as well as the PYTHIA 6~\cite{Sjostrand:2001yu} photoproduction events, are also used together with the H1 detector simulation in order to determine acceptance, efficiency and backgrounds. as well as to estimate systematic uncertainties associated with the measurement. DJANGOH and RAPGAP are interfaced to HERACLES~\cite{Spiesberger:1992vu,Kwiatkowski:1990es,Kwiatkowski:1990cx} to simulate the QED radiative effects. The generated events are passed through a detailed simulation of the H1 detector response based on the GEANT3 simulation program~\cite{Brun:1994aa} and are processed using the same reconstruction and analysis chain as used for the data. For the determination of the detector effects both the RAPGAP and DJANGOH predictions are studied.

\section{\label{subsec:dataselection} Event selection}

The data set used for this analysis was collected with the H1 detector in the years 2006 and 2007 when positrons and protons were collided at energies of 27.6\gev~and 920\gev, respectively. The integrated luminosity of the data set is $136$~\pbarnt~\cite{Aaron:2012kn}. DIS events were recorded using triggers based on electromagnetic energy deposits in the SpaCal
calorimeter. The trigger inefficiency is determined using independently triggered data and is negligible in the kinematic region of the analysis. 

The scattered positron is defined by the energy cluster in the SpaCal calorimeter with the highest transverse momentum. The energy of the cluster is further required to be larger than 12\gev, and the radial position of the cluster is required to be between 15 and 70 cm with respect to the beam axis. 
The $z$ coordinate of the event vertex is required to be within $|z_v|<35~\rm{cm}$ of the nominal interaction point. The $E-p_{z}$ variable is required to be between 35 and 75\gev~in order to reduce QED radiation and photoproduction backgrounds. Here $E-p_{z}$ is defined as the sum of $E_{i}-p_{z,i}$ of both the scattered positron and the hadronic final-state (HFS) particles. The HFS particles are reconstructed using an energy flow algorithm~\cite{Peez:2003zd,Hellwig:2004bb,Portheault:2005uu}.  This algorithm combines charged particle tracks and calorimetric energy clusters into hadronic objects, taking into account their respective resolution and geometric overlap, while avoiding double counting of energy. 
%

Tracks measured in the CTD alone (central tracks) and the combination of CTD and FTD (combined tracks) are used in this analysis. Both central and combined tracks are required to have transverse momenta $p_{\rm T, lab}>150$\mev. Furthermore, the total momentum for the combined tracks is required to be larger than 500\mev\ in order to ensure particles have enough momentum to cross the material between the CJC and FTD. The pseudorapidity of both central and combined tracks is required to be within $|\eta_{_{\text{lab}}}|<1.6$. In addition, the central tracks are required to have their $|dca'\sin{\theta}|<2~\rm{cm}$, where $dca'$ is the distance of closest approach with respect to the primary vertex and $\theta$ is the polar angle of the track. The innermost hit in the CTD is required for central tracks to be less than 50 cm away from the $z$ axis, where the radial track length is required to be larger than 10 cm. Neutral particles are not considered in the multiplicity analysis. Using only tracks assigned to the primary event vertex, the contributions from in-flight decays of $K^{0}_{S}$, $\Lambda$, from photon conversions and from other secondary decays and interactions with detector material are reduced. Details of the track selection are described elsewhere~\cite{Alexa:2013vkv}. 

\subsection{\label{subsec:eventreco} Event reconstruction}

The scattered positron and the HFS particles are used for the reconstruction of the kinematic variables, \xbj, $y$ and \qsqnospace. Similar to the H1 measurement of charged particle momentum spectra~\cite{Alexa:2013vkv}, the $e$-$\Sigma$ method is used~\cite{Bassler:1994uq}, where these variables are defined as

\be
\qsq=4E_{e}E^{'}_{e}\cos{\frac{\theta_{e}}{2}}^{2}, \;\;\;
y=2E_{e}\frac{\Sigma}{[\Sigma+E^{'}_{e}(1-\cos{\theta_{e}})]^{2}},\;\;\; 
x_{\rm bj}=\frac{\qsq}{sy}.
\ee

\noindent Here, $s$ is the $ep$ centre-of-mass energy squared, $E_e$ is the incoming lepton energy, $E^{'}_e$ and $\theta_{e}$ are the scattered positron energy and polar angle, respectively. The quantity $\Sigma$ is defined as $\sum{E_{i}-p_{z,i}}$, where the sum runs over all the HFS particles. This method provides an optimum in resolution of the kinematic variables and shows only little sensitivity to QED radiative effects. The hadronic centre-of-mass energy $W$ is defined as: 

\be
\label{eq:equation3}
W = \sqrt{sy-Q^{2}+M^{2}_{p}},
\ee

\noindent where $M_{p}$ is the proton rest mass.

The hadronic centre-of-mass frame is defined as the frame where $p+q=0$, with $p$ and $q$ being the four-momenta of the proton and the virtual photon, respectively. In this frame, the positive $z$ axis is aligned with the direction of the virtual photon.

\begin{figure}[tbh]
  \centering
\includegraphics[width=5.8in]{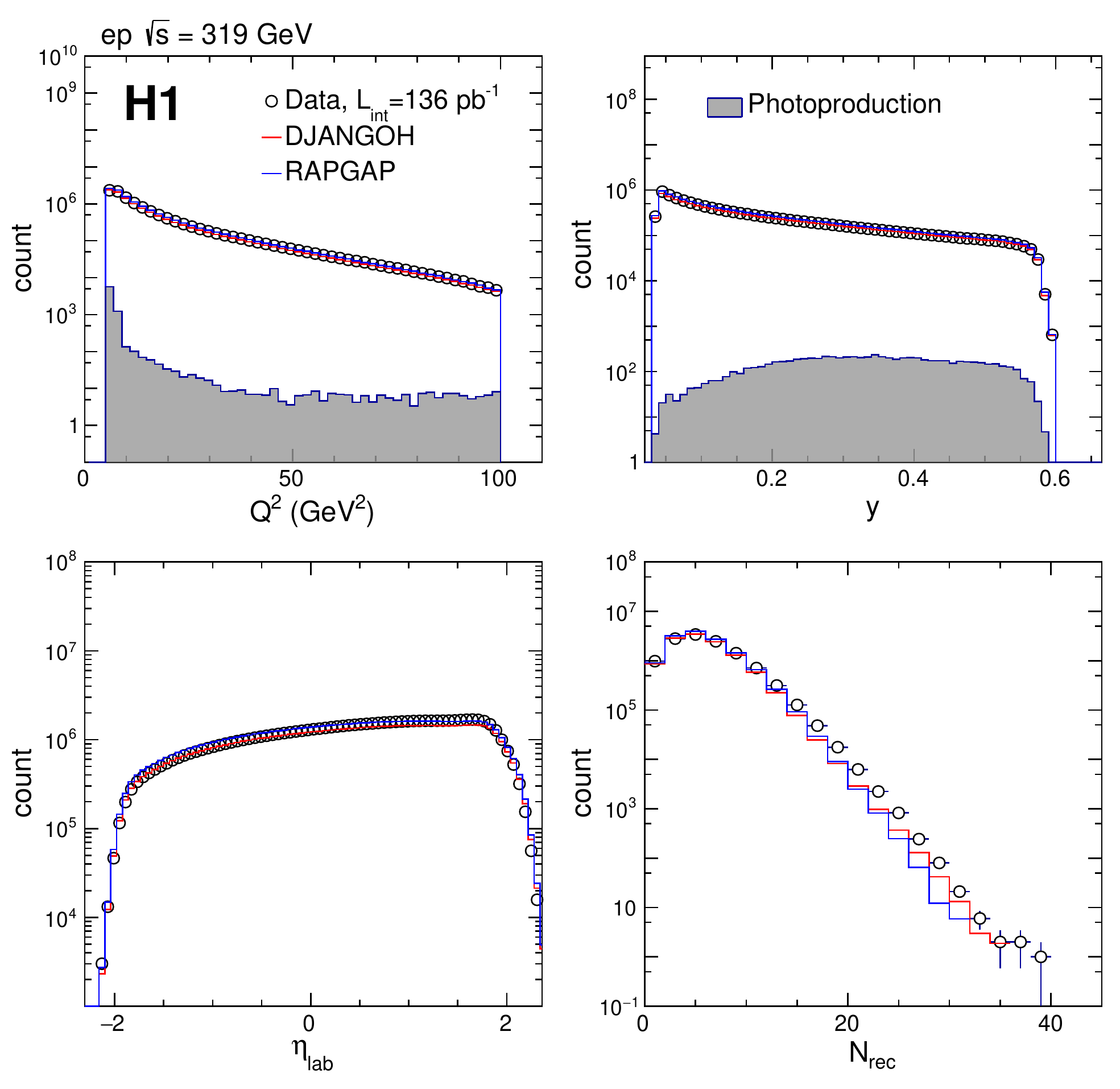
}
  \caption{ \label{fig:figure_control_1} Reconstructed momentum transfer \qsqnospace, inelasticity $y$, pseudorapidity $\eta_{_{\rm{lab}}}$, and charged particle multiplicity $N_{\rm rec}$ for data (open circle), and the DJANGOH (red line) and the RAPGAP (blue line) MC models. The phase space restrictions are given in Table~\ref{tab:phasespace}. The photoproduction background simulated from PYTHIA 6.4~\cite{Sjostrand:2006za} is shown for the \qsq and $y$ distributions. Error bars indicate the statistical uncertainty of data.}
\end{figure}

In figure~\ref{fig:figure_control_1}, distributions of the reconstructed quantities of \qsqnospace, $y$, $\eta_{_{\text{lab}}}$, and the number of charged particles $N_{\rm rec}$, are shown in comparison with predictions from DJANGOH and RAPGAP. The panels of the \qsq and $y$ distribution also contain the contribution of expected photoproduction background estimated based on PYTHIA 6.4~\cite{Sjostrand:2006za}. The photoproduction background is found to be less than 0.5\%, and therefore neglected in the subsequent analysis.

\subsection{\label{subsec:observable} Experimental observables and kinematic phase space}

For a given range in \qsq and $y$, the probability $P(N)$ is defined as the
fraction of events for which $N$ charged particles are produced in the
specified $\eta$ range relative to the total number of events in that \qsqnospace,$y$
range. Based on the distribution $P(N)$, the first and second moments of the multiplicity distributions, the KNO function $\Psi(z)$ and the final-state hadron entropy $S_{\rm hadron}$ are defined as:

\be
\langle N \rangle \equiv \frac{\sum{N\cdot P(N)}}{\sum{P(N)}},\;\;\;\;\;\;
Var(N) \equiv \frac{\sum{(N-\langle N \rangle)^{2}\cdot P(N)}}{\sum{P(N)}}, \\
\ee
\\
\be
\Psi(z)=\langle N \rangle P(N),
\ee
\\
\be
S_{\rm{hadron}} = -\sum{P(N)\ln{P(N)}}.
\ee

\noindent Here, the sum runs over the number of charged particles and the $z$ variable is equal to $N/\langle N \rangle$. These quantities are measured within the fiducial kinematic phase space listed in Table~\ref{tab:phasespace}. The selection in the HCM frame relative to the laboratory frame differs only by the additional restriction of the charged particles properties to the current hemisphere, $0<\eta^{\ast}<4$.

\begin{table}[h]
\begin{center}
\begin{tabular}{l|c|c}
\hline\hline
& laboratory frame & HCM frame\\\hline
$Q^2$&$5<Q^2<100\,\gevsq$&$5<@^2< 100\,\gevsq$\\
$y$&$0.0375 <y< 0.6$&$0.0375 <y< 0.6$\\
$p_{\text{T,lab}}$ &$p_{\text{T,lab}}>150\,\text{MeV}$&$p_{\text{T,lab}}>150\,\text{MeV}$\\
$\eta_{_{\text{lab}}}$&$-1.6 <\eta_{_{\text{lab}}}<1.6$&$-1.6 <\eta_{_{\text{lab}}}<1.6$\\
$\rap$ & -- & $0<\rap< 4$\\
\hline\hline
\end{tabular}
\caption{\label{tab:phasespace} Summary of the fiducial kinematic phase space used in this analysis.}
 \end{center}
 \end{table}

\subsection{\label{subsec:datacorrection} Data correction}

The detector-level charged track multiplicity distributions are corrected to stable particles with proper lifetime $c\tau>10~\rm{mm}$ including charged hyperons. 
The probabilities $P(N)$ are derived from the distributions of events with
$N=0,1,2,...$ tracks reconstructed in the specified $\eta$ range and with \qsq
and $y$ reconstructed in the kinematic bin of interest. First, the
observed event counts in the three-dimensional grid of $N$, \qsqnospace,
$y$ are unfolded, such that migrations between bins as well as
efficiency and acceptance distortions are removed. The second step is to
normalize the event counts as a function of $N$ to the total number
of events in each \qsqnospace, $y$ bin and thus obtain the probabilities $P(N)$.
The last step is to correct for QED radiation from the electron line.

The unfolding is done within the TUnfold framework~\cite{Schmitt:2012kp}. In order to better
resolve migration effects, the number of bins in \qsqnospace, $y$ and $N$ is
chosen to be higher when counting events in reconstructed quantities as 
compared to the truth quantities. The phase space in \qsq and $y$ is enlarged 
over the phase space given in Table~\ref{tab:phasespace}, such that the measured phase space
is guarded against migrations at the phase space boundaries. 
The unfolding turns out to be robust against variations of the
regularisation scheme. The dominant systematic uncertainty in the
detector correction procedure arises from constructing the matrix of
migration with an alternative MC model, DJANGOH instead of the RAPGAP
default.

In this paragraph QED radiative effects are discussed. In the radiative MC, the scattered positron at the particle level is defined to be a scattered positron with photons that are within a cone of 0.4 radian. The QED radiative effects are corrected for based on the MC event generator. The correction factors are derived bin-by-bin in each measured phase space at the particle level using radiative and non-radiative MC samples. 

The binning of the multiplicity $N$ at the particle level after the unfolding is made to be wider than at the detector level when $N>3$, in order to avoid large negative bin-to-bin correlations. For those wide bins, the reported values of $P(N)$ are defined as $\sum{P(N)/\Delta}$, where $\Delta$ is the number of distinct values of $N$ included in the bin. In order to determine moments and the hadron entropy, the measured distribution is extrapolated within the wide bins using an exponentiated cubic spline. Correlations between the extrapolated values are taken into account. Bin centres of the wide bins are also reported.

\section{Systematic uncertainties}

Systematic uncertainties are studied based on systematic variations on fully corrected results, where details of the variations are listed below. The systematic uncertainties are found to depend on $N$, $\eta_{_{\text{lab}}}$, \qsq and $y$.
They are are observed to be similar in size for the HCM results as compared to those obtained in the laboratory frame.
The following systematic uncertainty sources are considered in this analysis: 

\begin{itemize}
\item Sys.1 - \textbf{Radiative corrections}: the difference in correction factors between the MC generators, the DJANGOH and RAPGAP is taken as systematic uncertainty. It is found to be 1--2\% for the $P(N)$ distributions and up to 1\% for their moments. 
\item Sys.2 - \textbf{MC model dependence}: the $P(N)$ distributions and their moments are compared between results that are corrected by DJANGOH or RAPGAP. This leads to 1--4\% systematic uncertainty on the multiplicity distributions and their moments.
\item Sys.3 - \textbf{SpaCal energy scale and angular resolution}: a variation of 1\% on the energy scale of the SpaCal\cite{Aaron:2009bp} and $1^{\circ}$ on the angular direction is considered. The combined systematic uncertainty on the distribution of $P(N)$ and its moments is found to be 1--5\% and 2\%, respectively.
\item Sys.4 - \textbf{Hadronic energy scale}: a variation of 2\% on the energy scale of hadronic final-state objects~\cite{Salek:2010ifa} results in 1--3\% systematic uncertainty on the multiplicity distribution and 2\% on the moments.
\item Sys.5 - \textbf{Single track efficiency}: 0.5\% of tracking efficiency uncertainty on central tracks and 10\% uncertainty on combined tracks are applied. This leads to 1--5\% systematic uncertainty on the $P(N)$ distributions, and 1\% on the moments.
\item Sys.6 - \textbf{V0 particle contamination}: 50\% of uncertainty on tracks originating from $K^{0}_{s}$ decay and 0.5\% uncertainty on tracks from photon conversion and Dalitz-decays surviving the primary track selection result in 1--7\% of systematic uncertainty on the $P(N)$ distributions and up to 2\% for the moments.
\item Sys.7 - \textbf{Diffractive contributions}: variation of the diffractive MC contribution by a factor of 2 results in 1--5\% systematic uncertainty on the $P(N)$ distributions and up to 1\% on the moments.
\item Sys.8 - \textbf{Extrapolation}: values of $P(N)$ that are not explicitly measured are extrapolated and 1\% of difference is found for the mean, variance, and entropy with respect to the generator value based on MC. 
\end{itemize} 

\noindent Systematic uncertainties associated with photoproduction background are negligible. A summary of systematic uncertainties as a function of the multiplicity $N$ can be found in Table~\ref{tab:systematics}. The total systematic uncertainties are obtained by adding in quadrature all individual contributions. The tables of the results contain a complete breakdown of the systematic uncertainty contributions from different sources for each measured data point.

\begin{table}[h]
\fontsize{10}{15}\selectfont
\begin{center}
\begin{tabular}{lcccccccc}
\hline
\hline
\noalign{\vskip 1mm}
  \textbf{Systematic sources} & $\rm N(0,1)$ & $\rm N(2,4)$  & $\rm N(5,14)$  & $\rm N(15,\inf) $ & $\langle N \rangle$ & $Var(N)$  & $S_{\rm{hadron}}$ \\
  \noalign{\vskip 1mm}
\hline
\noalign{\vskip 1mm}
 Radiative corrections (\%) & 1.8 & 1.0 & 1.4 & 1.8 & 1.0 & 1.0 & 0.0 \\
 MC model dependence (\%) &  4.1 &  2.2  & 1.8  & 3.9 & 1.5 & 4.3 & 1.3 \\
 SpaCal energy and polar angle (\%) & 5.0 & 1.0  &  1.0 &  2.0 & 2.0 & 2.0  & 1.0 \\
 Hadronic energy scale (\%) & 3.0 & 1.0  & 1.0 & 3.0 & 2.0 & 2.0 & 1.0 \\
 Single track efficiency (\%) & 1.0  & 1.0   & 5.0 & 5.0  & 1.0 & 1.0 & 0.5 \\
 V0s particle contamination (\%) & 1.0  & 2.0   & 5.0 & 7.1  & 1.5 & 2.0 & 0.5 \\
 Diffractive contributions  (\%) & 5.8  & 1.0   & 1.1 & 1.0  & 0.5 & 1.0 & 0.0 \\
 Extrapolation (\%) & -  & -   & - & -  & 1.0 & 1.0 & 1.0 \\
 \noalign{\vskip 1mm}
\hline
\noalign{\vskip 1mm}
 \textbf{Total uncertainty} (\%) & 9.8  &  3.8  &  7.6 & 10.8 & 4.0 & 5.9 & 2.3\\
 \noalign{\vskip 1mm}
\hline
\hline
 \end{tabular}
 \end{center}
  \caption{\label{tab:systematics} Summary of the systematic uncertainties in this analysis.}
 \end{table}

\section{\label{sec:result} Results}
\subsection{\label{subsec:result_multiplicity} Multiplicity distributions}

The charged particle multiplicity distributions in $ep$ DIS at $\sqrt{s}=319$\gev~are measured in different bins of \qsq and $y$ for particles with $p_{_{\rm T,lab}}>150~\mev$ and $|\eta_{_{\rm{lab}}}|<1.6$ in the laboratory frame. The data are presented in figure~\ref{fig:figure_5_1}, numerical values are reported in the appendix. The data are compared with predictions from the DJANGOH, RAPGAP, and PYTHIA 8 event generators without simulation of QED radiation. The multiplicity distributions $P(N)$ are found to broaden as $y$ increases for fixed \qsqnospace. Since $y$ can be related to the hadronic centre-of-mass energy $W$ (cf. Eq.~\ref{eq:equation3}), the increase in multiplicity is qualitatively expected because more energy is available for hadronisation. On the other hand, the $P(N)$ distributions are found to be almost independent of \qsq for fixed $y$. Qualitatively, the MC predictions can describe the peak position of the multiplicity distributions well. However, they tend to underestimate the data both at low and high multiplicities, especially towards low \qsq and high $y$. Among all MC models considered, the PYTHIA 8 model gives the poorest description and peaks significantly below the data, especially at high $y$.

\begin{figure}[tbh]
  \centering
\includegraphics[width=5.8in]{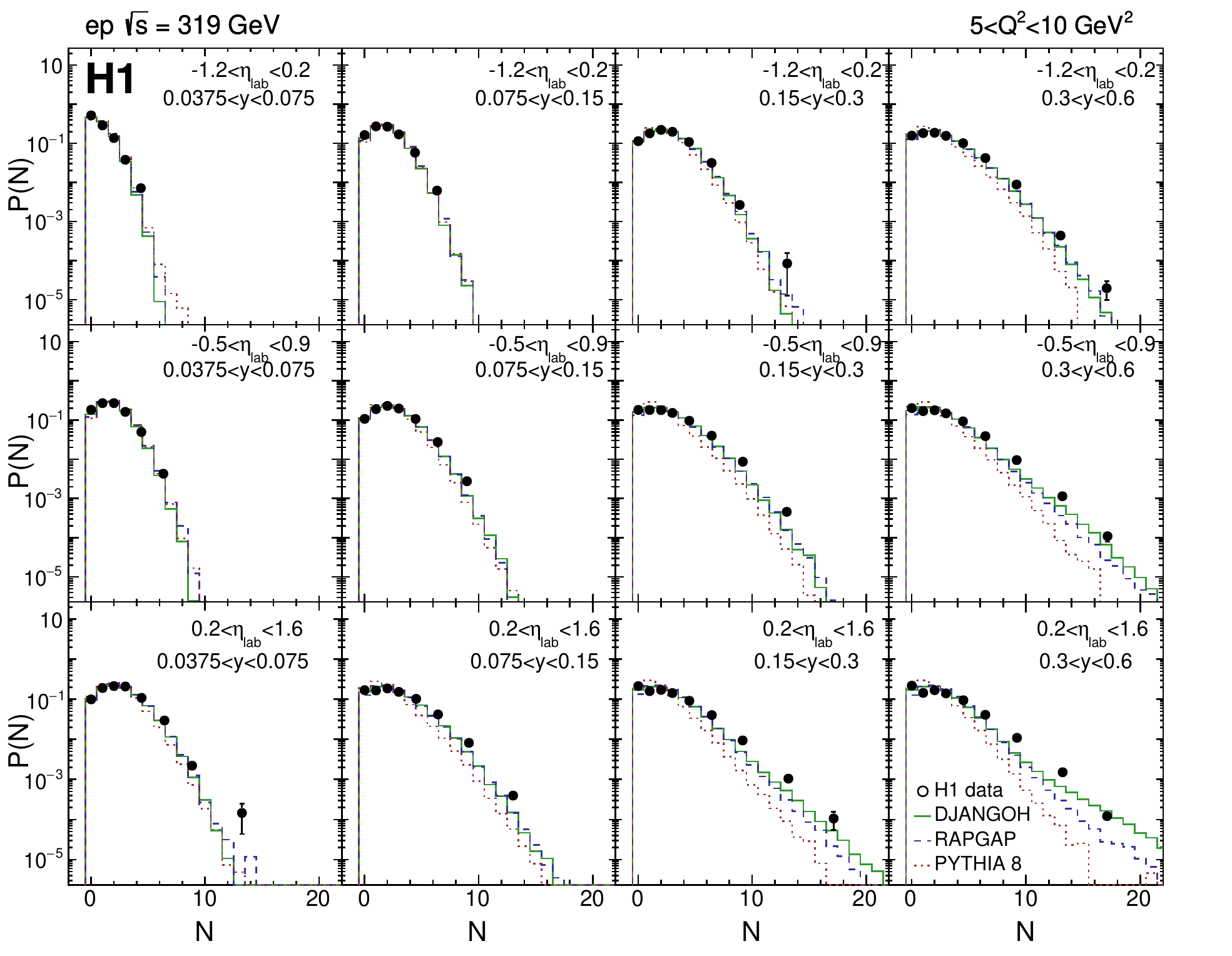
}
  \caption{ \label{fig:figure_5_2} 
  Charged particle multiplicity distributions $P(N)$ as a function of the number of particles $N$ at $\sqrt{s}=319$\gev~ep collisions in the range $5<\qsq<10~\gevsq$. Further phase space restrictions are given in Table~\ref{tab:phasespace}. Different panels correspond to different $\eta_{_{\text{lab}}}$ and $y$ bins, as indicated by the text in the figure. Predictions from DJANGOH, RAPGAP and PYTHIA 8 are also shown. The total uncertainty is denoted by the error bars.}
\end{figure}

\begin{figure}[tbh]
  \centering
\includegraphics[width=5.8in]{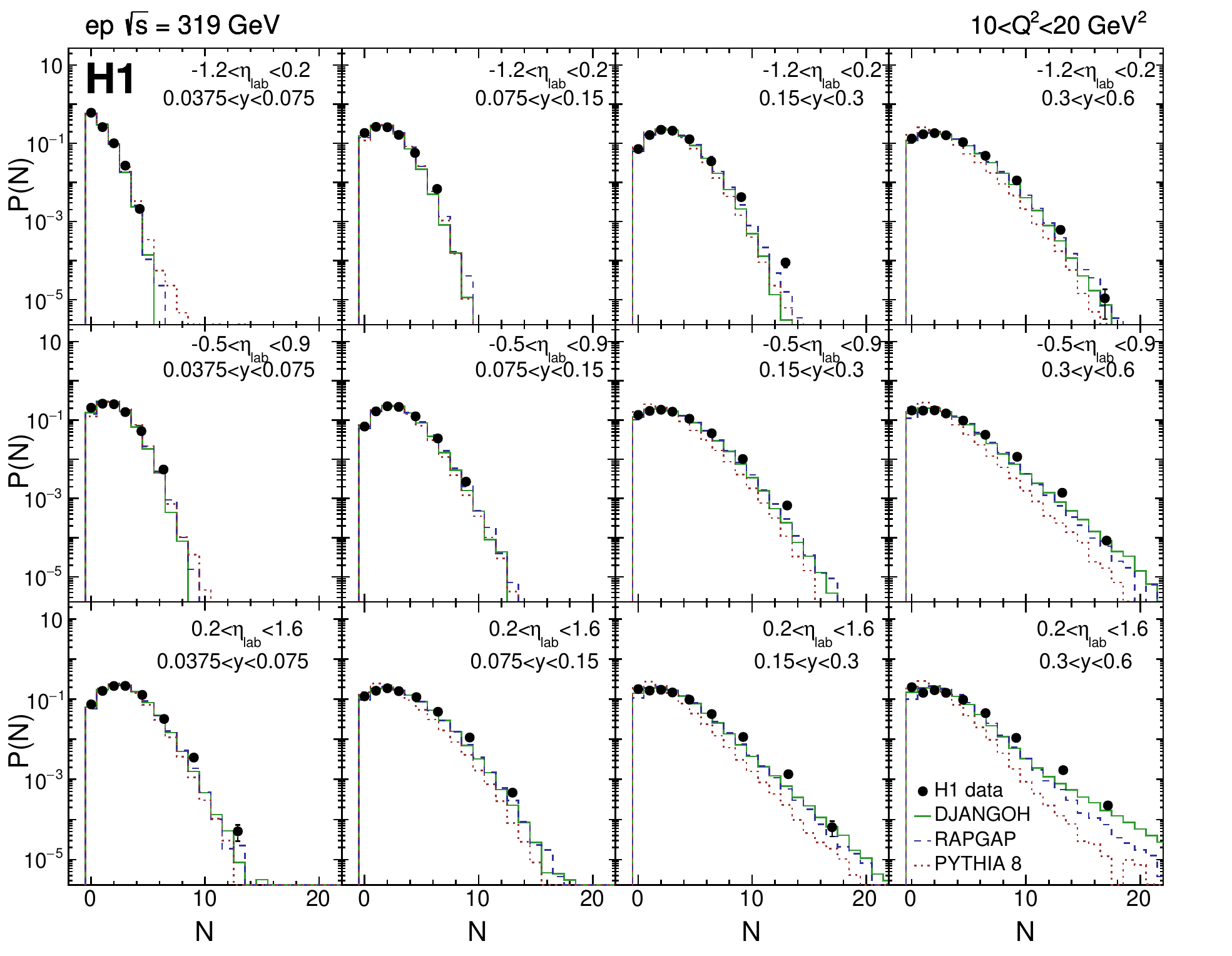
}
  \caption{ \label{fig:figure_5_3} 
  Charged particle multiplicity distributions $P(N)$ as a function of the number of particles $N$ at $\sqrt{s}=319$\gev~ep collisions in the range $10<\qsq<20~\gevsq$. Further phase space restrictions are given in Table~\ref{tab:phasespace}. Different panels correspond to different $\eta_{_{\text{lab}}}$ and $y$ bins, as indicated by the text in the figure. Predictions from DJANGOH, RAPGAP and PYTHIA 8 are also shown. The total uncertainty is denoted by the error bars.}
\end{figure}

\begin{figure}[tbh]
  \centering
\includegraphics[width=5.8in]{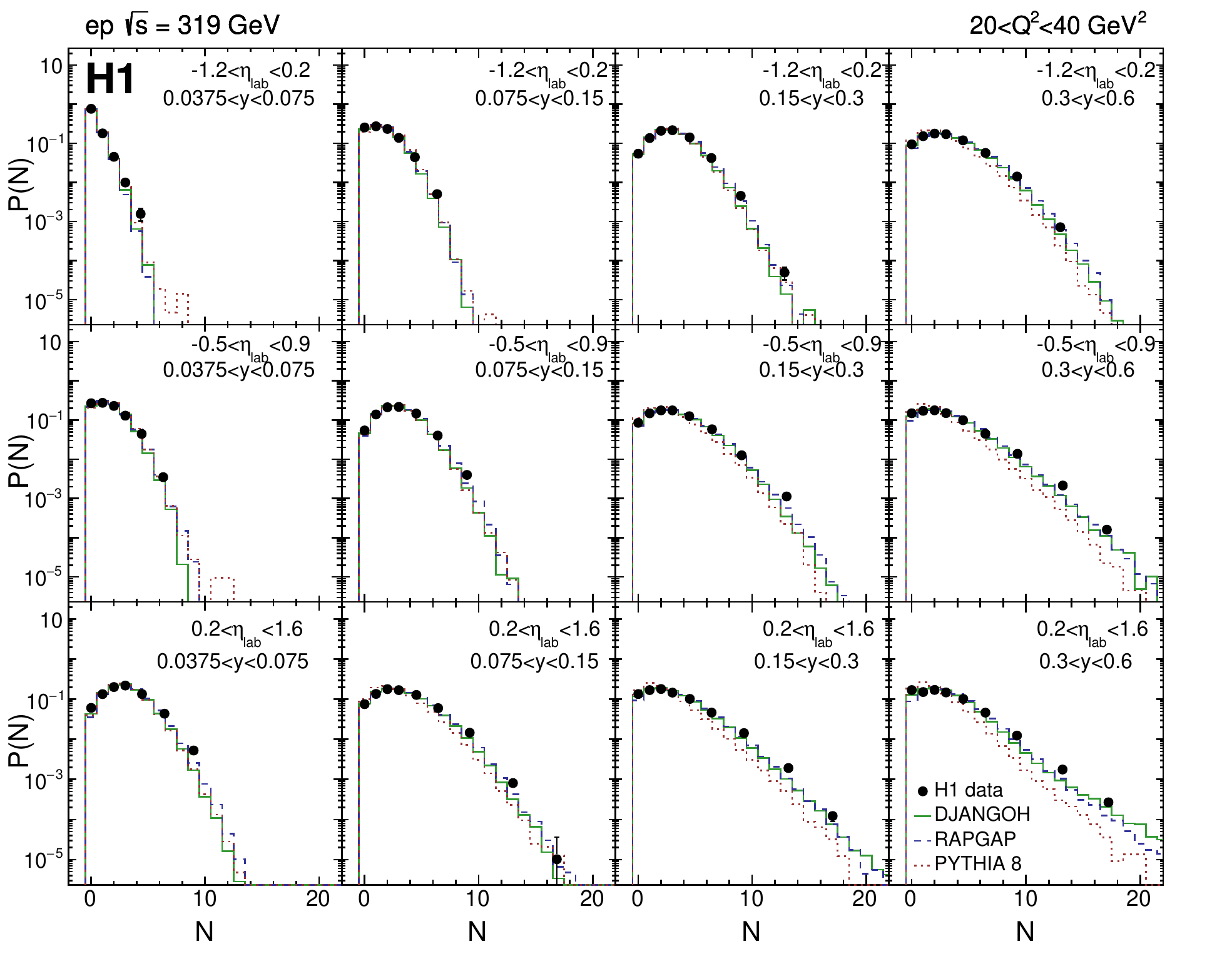
}
  \caption{ \label{fig:figure_5_4} 
  Charged particle multiplicity distributions $P(N)$ as a function of the number of particles $N$ at $\sqrt{s}=319$\gev~ep collisions in the range $20<\qsq<40~\gevsq$. Further phase space restrictions are given in Table~\ref{tab:phasespace}. Different panels correspond to different $\eta_{_{\text{lab}}}$ and $y$ bins, as indicated by the text in the figure. Predictions from DJANGOH, RAPGAP and PYTHIA 8 are also shown. The total uncertainty is denoted by the error bars.}
\end{figure}

\begin{figure}[tbh]
  \centering
\includegraphics[width=5.8in]{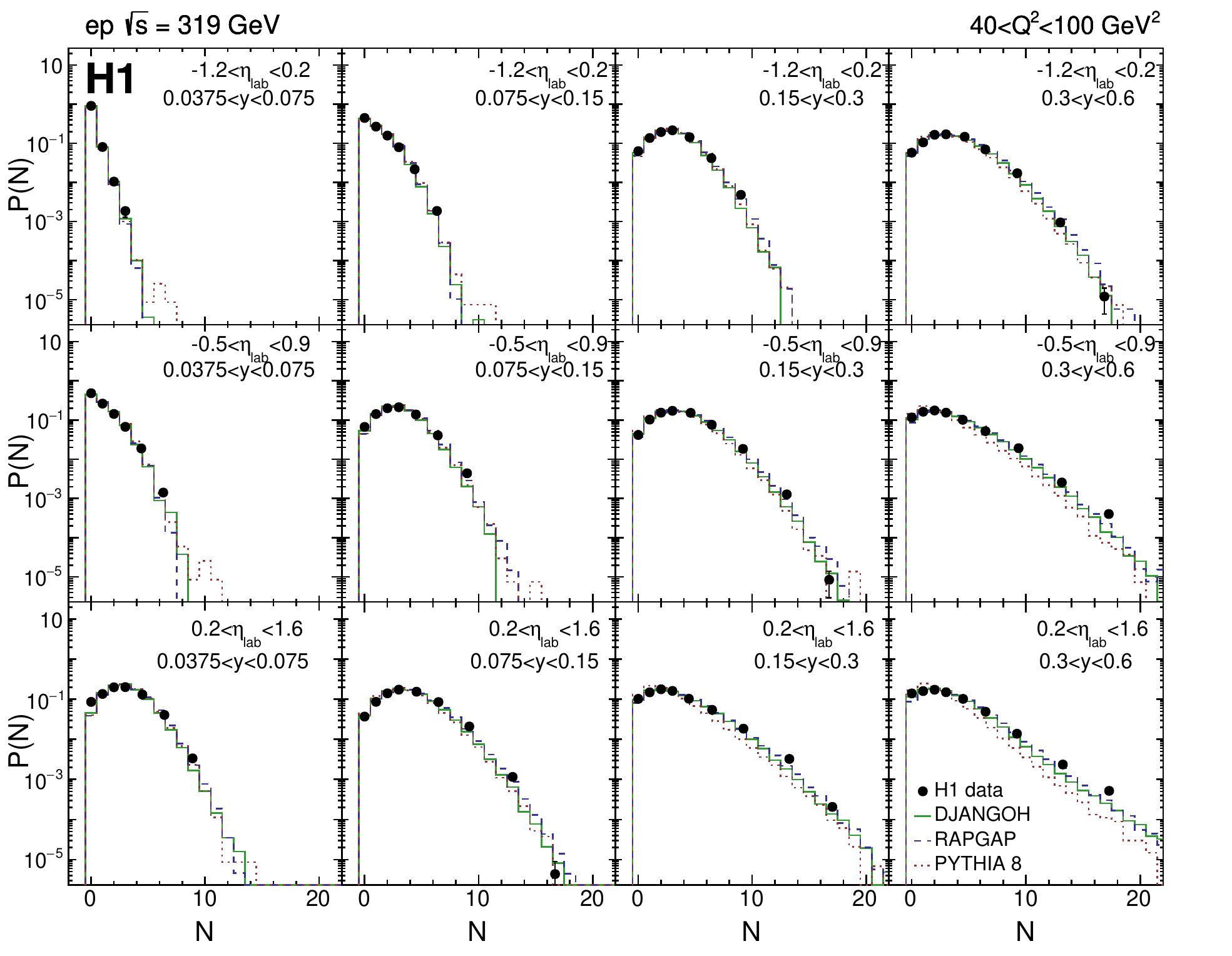
}
  \caption{ \label{fig:figure_5_5} 
  Charged particle multiplicity distributions $P(N)$ as a function of the number of particles $N$ at $\sqrt{s}=319$\gev~ep collisions in the range $40<\qsq<100~\gevsq$. Further phase space restrictions are given in Table~\ref{tab:phasespace}. Different panels correspond to different $\eta_{_{\text{lab}}}$ and $y$ bins, as indicated by the text in the figure. Predictions from DJANGOH, RAPGAP and PYTHIA 8 are also shown. The total uncertainty is denoted by the error bars.}
\end{figure}

In Figures.~\ref{fig:figure_5_2} to~\ref{fig:figure_5_5} the charged particle multiplicity distributions $P(N)$ are presented in four bins of \qsqnospace. In each Figure, the $P(N)$ distributions are shown differentially in bins of $y$ (identical binning as in figure~\ref{fig:figure_5_1}) and in three different ranges of $\eta_{_{\text{lab}}}$. Numerical values are given in the appendix. The overlapping $\eta_{_{\text{lab}}}$ ranges are chosen such that a pseudorapidity window of 1.4 units around the scattered parton direction in the leading order Quark Parton Model (QPM) picture can be selected. Similar to the measurements over the full $\eta_{_{\text{lab}}}$ range, the $P(N)$ distributions are found to broaden as $y$ increases, independent of the considered $\eta_{_{\text{lab}}}$ ranges. For fixed $y$, the $P(N)$ distributions also broaden as $\eta_{_{\text{lab}}}$ increases. Similar to the results presented in figure~\ref{fig:figure_5_1}, the MC models underestimate the high multiplicity tail, where the deviation is found to be the strongest at low \qsqnospace.

\begin{figure}[tbh]
  \centering
\includegraphics[width=5.8in]{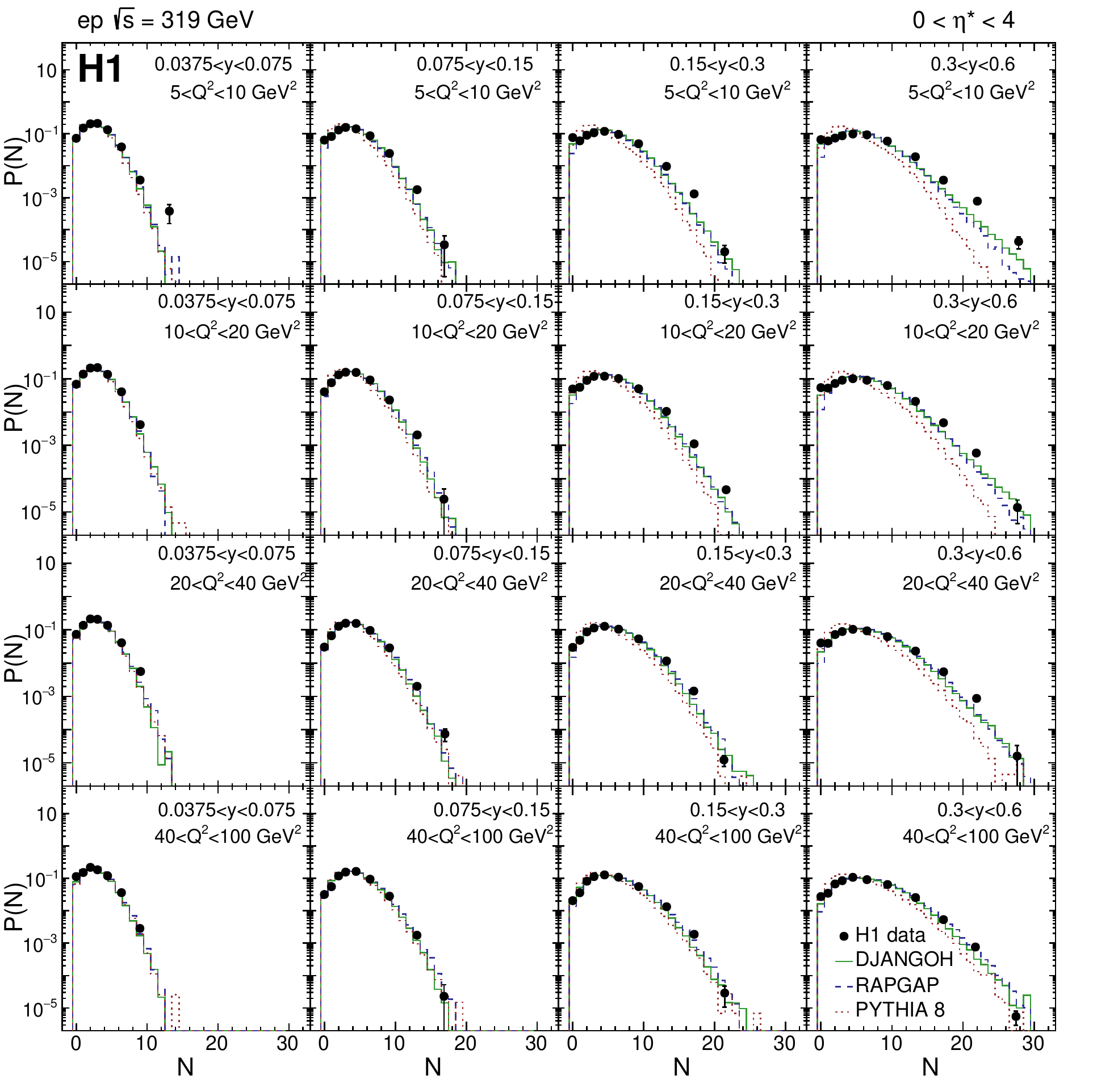
}
  \caption{ \label{fig:figure_5_10} 
  Charged particle multiplicity distributions $P(N)$ as a function of the number of particles $N$ at $\sqrt{s}=319$\gev~ep collisions with additional restriction to the current hemisphere $0<\eta^{\ast}<4$. Further phase space restrictions are given in Table~\ref{tab:phasespace}. Different panels correspond to different \qsq and $y$ bins, as indicated by the text in the figure. Predictions from DJANGOH, RAPGAP and PYTHIA 8 are also shown. The total uncertainty is denoted by the error bars. }
\end{figure}

In figure~\ref{fig:figure_5_10}, the multiplicity distributions $P(N)$ are presented with the additional restriction in the HCM frame $0<\eta^{\ast}<4$, mainly selecting particles originating from the current hemisphere. Numerical values are given in the appendix. Predictions obtained from DJANGOH, RAPGAP, and PYTHIA 8 are compared with data. Qualitatively, the comparisons are very similar to those of Figure.~\ref{fig:figure_5_1}. For the results presented in the subsequent sections, only RAPGAP is compared to the data.

\subsection{\label{subsec:result_moments} Moments of multiplicity distributions}

\begin{figure}[tbh]
  \centering
\includegraphics[width=3.0in]{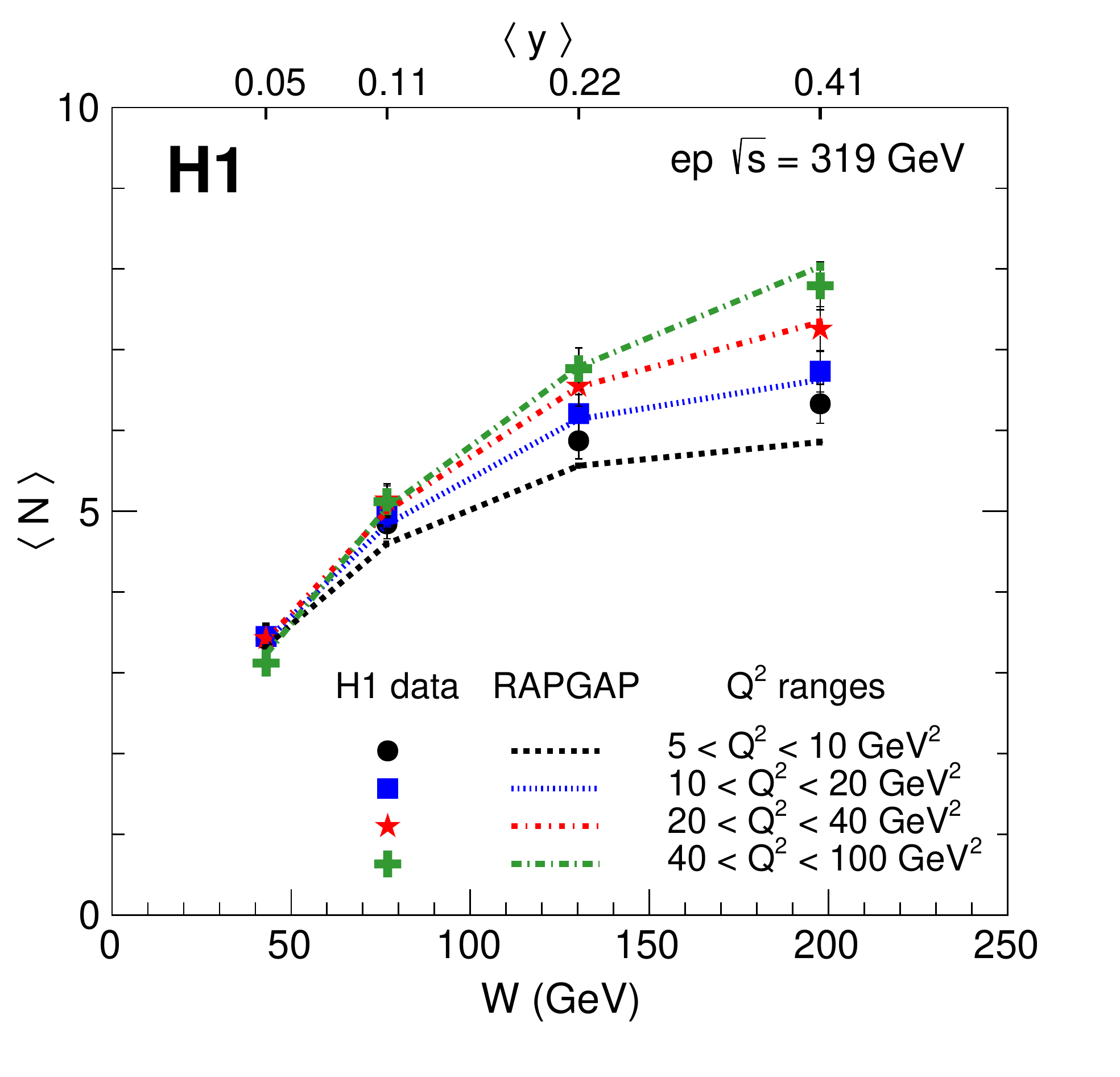}%
\includegraphics[width=3.0in]{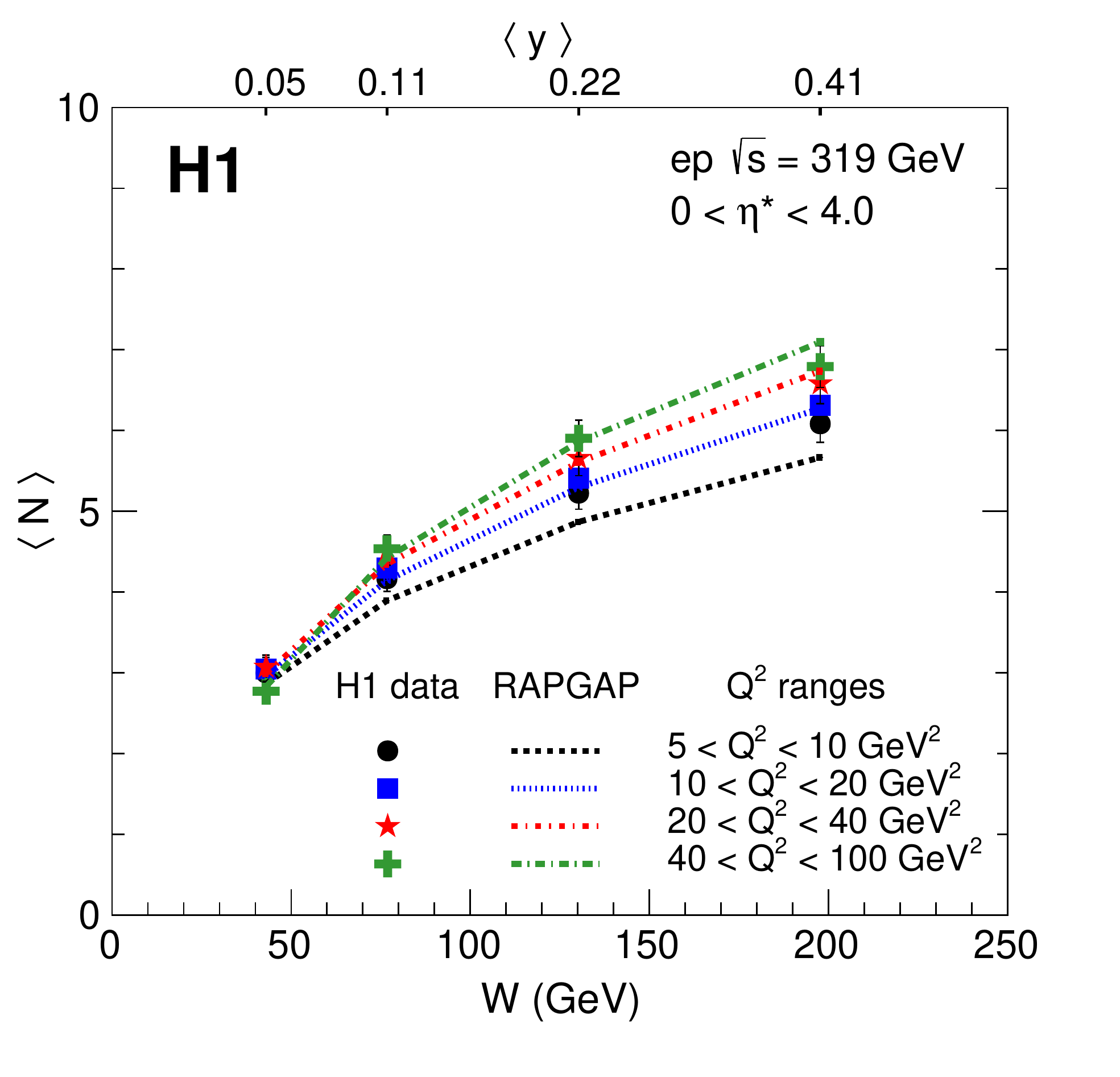}
  \caption{ \label{fig:figure_5_6} Mean multiplicity $\langle N \rangle$ as a function of $W$ measured at $\sqrt{s}=319$\gev~$ep$ collisions (left) and with an additional restriction to the current hemisphere $0<\eta^{\ast}<4$ (right). Further phase space restrictions are given in Table~\ref{tab:phasespace}. The corresponding $\langle y \rangle$ is indicated by the top axis for each measured $W$. Predictions from the RAPGAP model are shown by dashed lines. The total uncertainty is represented by the error bar. }
\end{figure}

The mean multiplicity $\langle N \rangle$ as a function of $W$ is presented in figure~\ref{fig:figure_5_6} and tables~\ref{tab:table_figure_8_a} to~\ref{tab:table_figure_8_b}, for both the full phase space and with an additional restriction in $\eta^{\ast}$ (see Table~\ref{tab:phasespace}). Predictions from the RAPGAP model are also shown. The average multiplicity $\langle N \rangle$ rises with the hadronic centre-of-mass energy $W$, which is in agreement with previous observations~\cite{Aad:2016xww,ALICE:2017pcy,Khachatryan:2010nk,Aaij:2014pza,Chekanov:2008ae,Adloff:1997fr,Aid:1996cb}. For higher \qsqnospace, the increase in $W$ is faster than that at lower \qsqnospace. For high $W$, the \qsq dependence is observed to be stronger in the full phase space than with the $\eta^{\ast}$ restriction of the current hemisphere. The RAPGAP prediction yields a reasonable description of the data. Only at high $W$ and low \qsq the MC tends to underestimate the data. 

\begin{figure}[tbh]
  \centering
\includegraphics[width=3.0in]{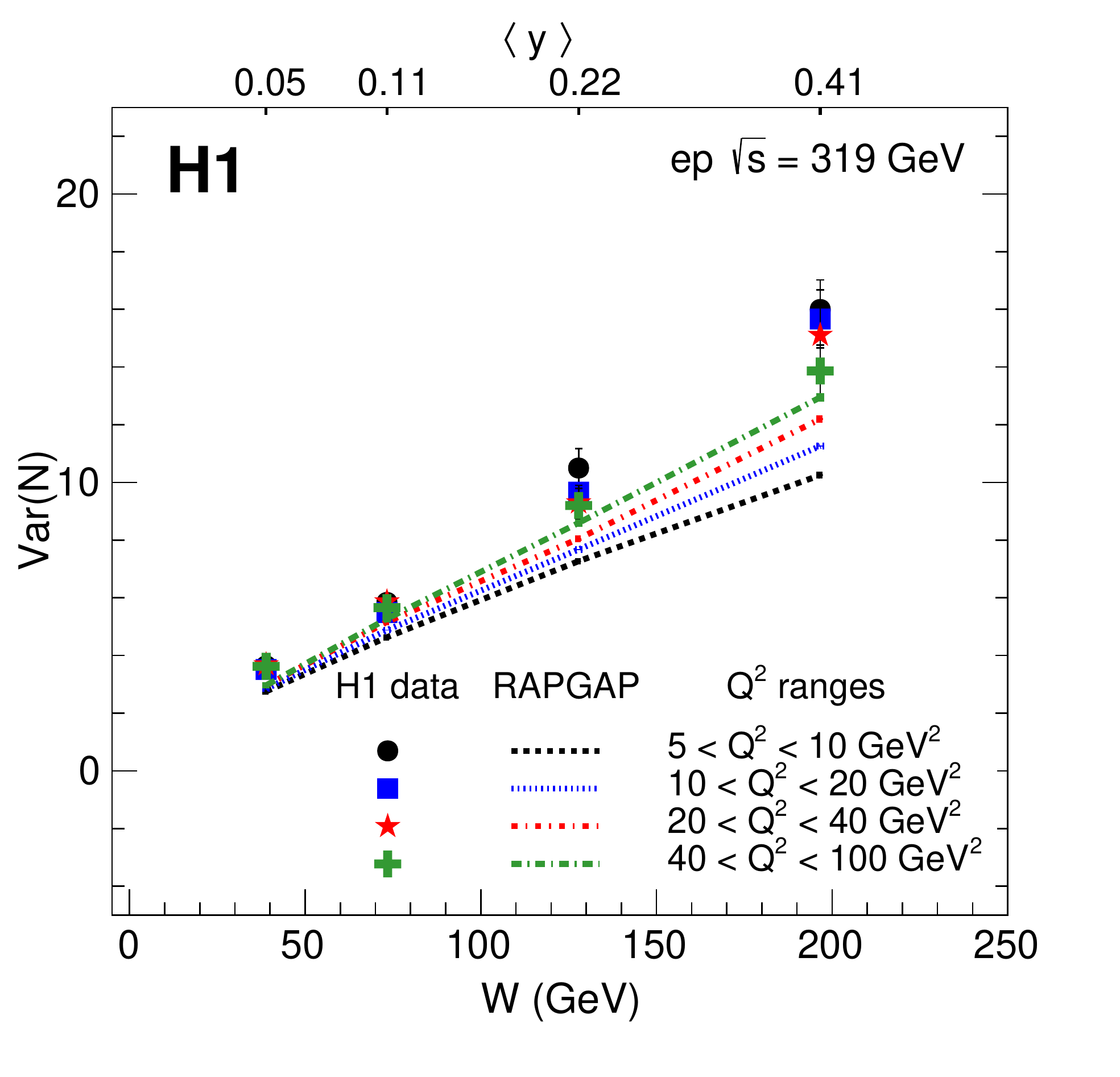}%
\includegraphics[width=3.0in]{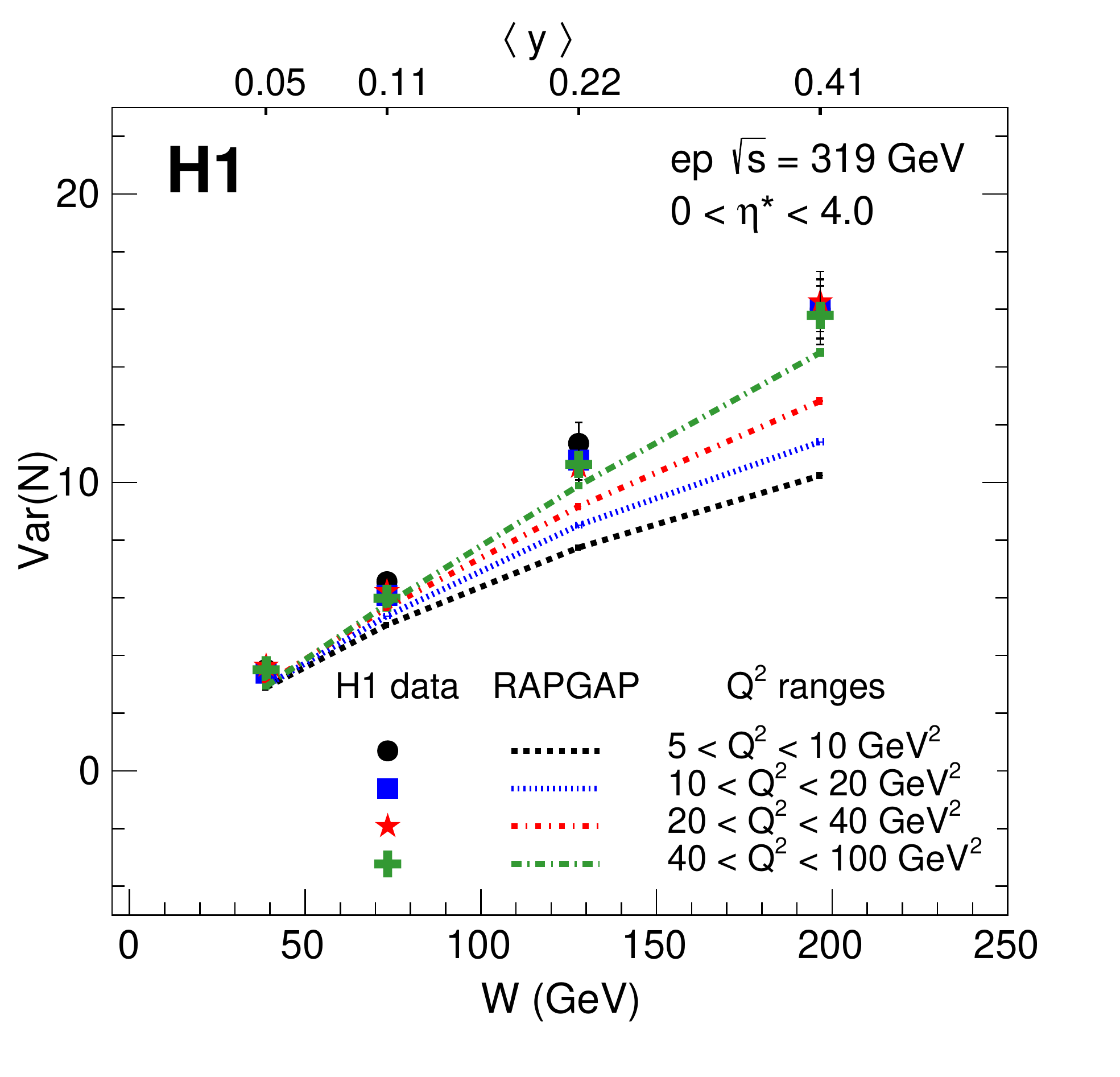}
  \caption{ \label{fig:figure_5_8} Second moment (variance) of the multiplicity distributions $Var(N)$ as a function of $W$ measured at $\sqrt{s}=319$\gev~$ep$ collisions (left) and with an additional restriction to the current hemisphere $0<\eta^{\ast}<4$ (right). Further phase space restrictions are given in Table~\ref{tab:phasespace}. The corresponding $\langle y \rangle$ is indicated by the top axis for each measured $W$. Predictions from the RAPGAP model are shown by dashed lines. The total uncertainty is represented by the error bar. }
\end{figure}

Similarly, in figure~\ref{fig:figure_5_8} and tables~\ref{tab:table_figure_9_a} to~\ref{tab:table_figure_9_b}, second moments of the multiplicity distributions (variances) are presented as a function of $W$, for both the full phase space and with the additional restriction in $\eta^{\ast}$. The variances rise strongly with $W$, almost independent of \qsq within uncertainties. The restriction of the current hemisphere has little inpact on the variance. For $\qsq>40~\gevsq$, RAPGAP describes the data reasonably well. However, towords high $W$, the MC not only underestimates the data but also shows a \qsq dependence, which is absent in data. This effect is more pronounced when restricting the analysis to the current hemisphere. 

\subsection{\boldmath The KNO-scaling function $\Psi(z)$}

In order to further study the multiplicity distribution, the KNO function $\Psi(z)$ measured as a function of $z=N/\langle N \rangle$ is shown in different bins of \qsq in figure~\ref{fig:figure_5_10_KNO}. Numerical values are given in the appendix. The analysis is restricted to the current hemisphere ($0<\eta^{\ast}<4$). In all \qsq bins, KNO scaling is observed, in broad agreement with many past experiments. However, in proton-proton collisions at LHC energies, violations of KNO scaling have been reported recently~\cite{Khachatryan:2010nk}.

\subsection{\label{subsec:result_entropy} Entropy}

\begin{figure}[tbh]
  \centering
\includegraphics[width=5.8in]{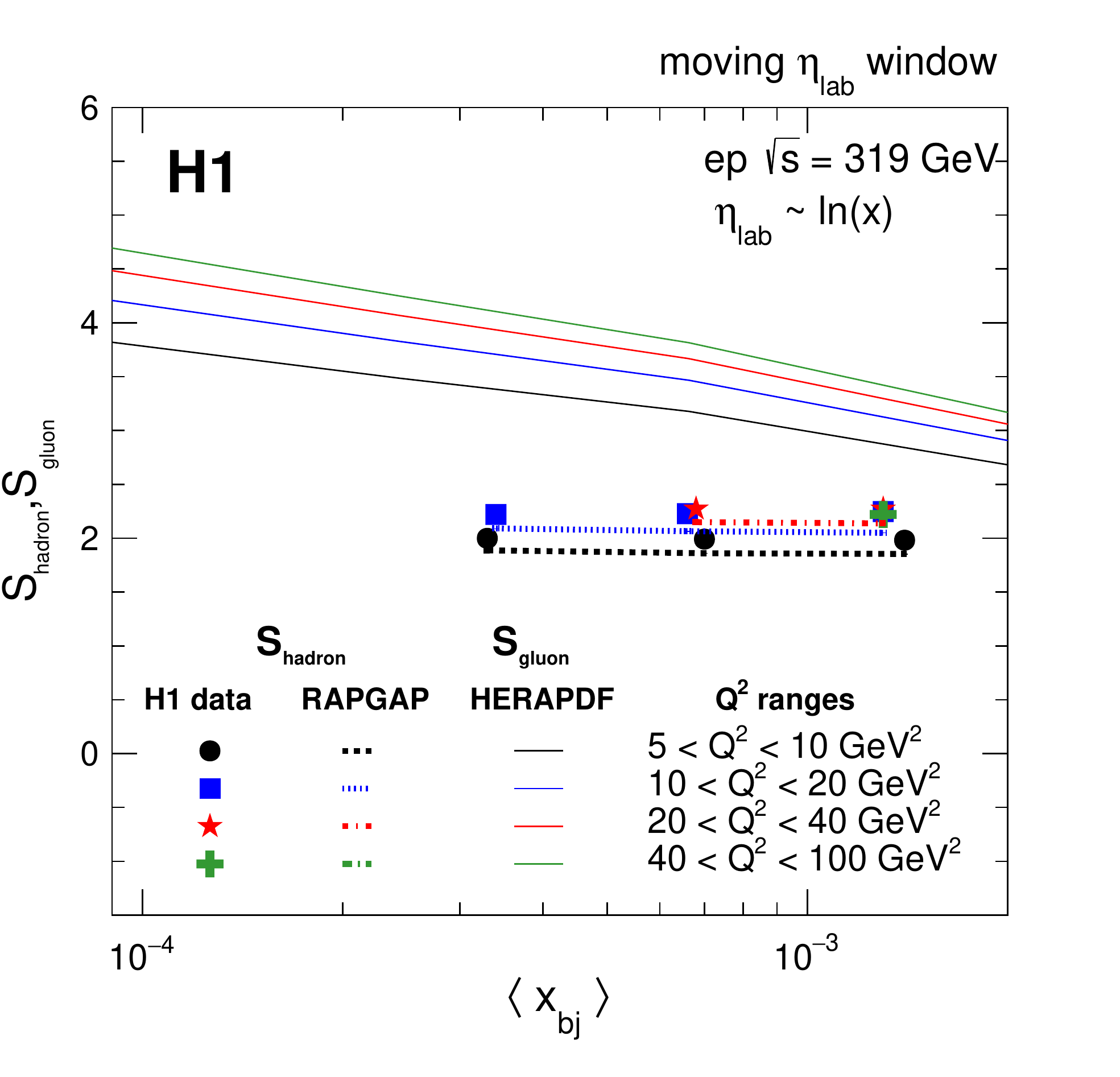
}
\caption{ \label{fig:figure_5_14} Hadron entropy $S_{\rm hadron}$ derived from  multiplicity distributions, reported as a function of $\langle$\xbj$\rangle$ in different \qsq ranges, measured in $\sqrt{s}=319$\gev~$ep$ collisions. For each $\langle$\xbj$\rangle$, the multiplicity is determined in a dedicated pseudorapidity window as discussed in the text. Further phase space restrictions are given in Table~\ref{tab:phasespace}.
  Predictions for $S_{\rm hadron}$ from the RAPGAP model and for the entanglement entropy $S_{\rm gluon}$ based on an entanglement model are shown by the dashed lines and solid lines, respectively. For each \qsq range, the value of the lower boundary is used for predicting $S_{\rm gluon}$.
  The total uncertainty on the data is represented by the error bars. }
\end{figure}

It was recently suggested~\cite{Kharzeev:2017qzs,Tu:2019ouv} that the final-state hadron entropy $S_{\rm hadron}$ calculated from charged particle multiplicity distributions might be related to the entanglement entropy of gluons $S_{\rm gluon}$ at low \xbj~(Eq.~\ref{eq1}). In figure~\ref{fig:figure_5_14} and Table~\ref{tab:table_figure_10}, $S_{\rm hadron}$ is studied as a function of $\langle$\xbj$\rangle$ in different \qsq bins.
Similar to the observable studied in Ref.~\cite{Tu:2019ouv}, a moving $\eta_{_{\rm{lab}}}$ window depending on $\langle$\xbj$\rangle$ is chosen to match the rapidity of the scattered quark in a leading order QPM picture.
The respective selected pseudorapidity window in the laboratory frame is indicated in Table~\ref{tab:table_figure_10}.
The hadron entropy observed in data is cosistent with being constant in $\langle$\xbj$\rangle$, increasing only slightly with \qsqnospace. The same $\langle$\xbj$\rangle$ and \qsq dependence is also observed for RAPGAP, however with slightly smaller values of $S_{\rm hadron}$. Predictions for the entanglement entropy $S_{\rm gluon}$ based on the gluon density $xG(x,Q^{2})$ are also shown at various \qsq values, which correspond to the lower boundaries of the \qsq bins in data. Here the PDF set HERAPDF 2.0~\cite{Abramowicz:2015mha} at leading order is used. Neither the dependence on $\langle$\xbj$\rangle$ nor the magnitude agrees with data. Thus the prediction $S_{\rm hadron}=S_{\rm gluon}$ (Eq.~\ref{eq1}) is not confirmed by the present measurement.

\begin{figure}[tbh]
  \centering
\includegraphics[width=5.8in]{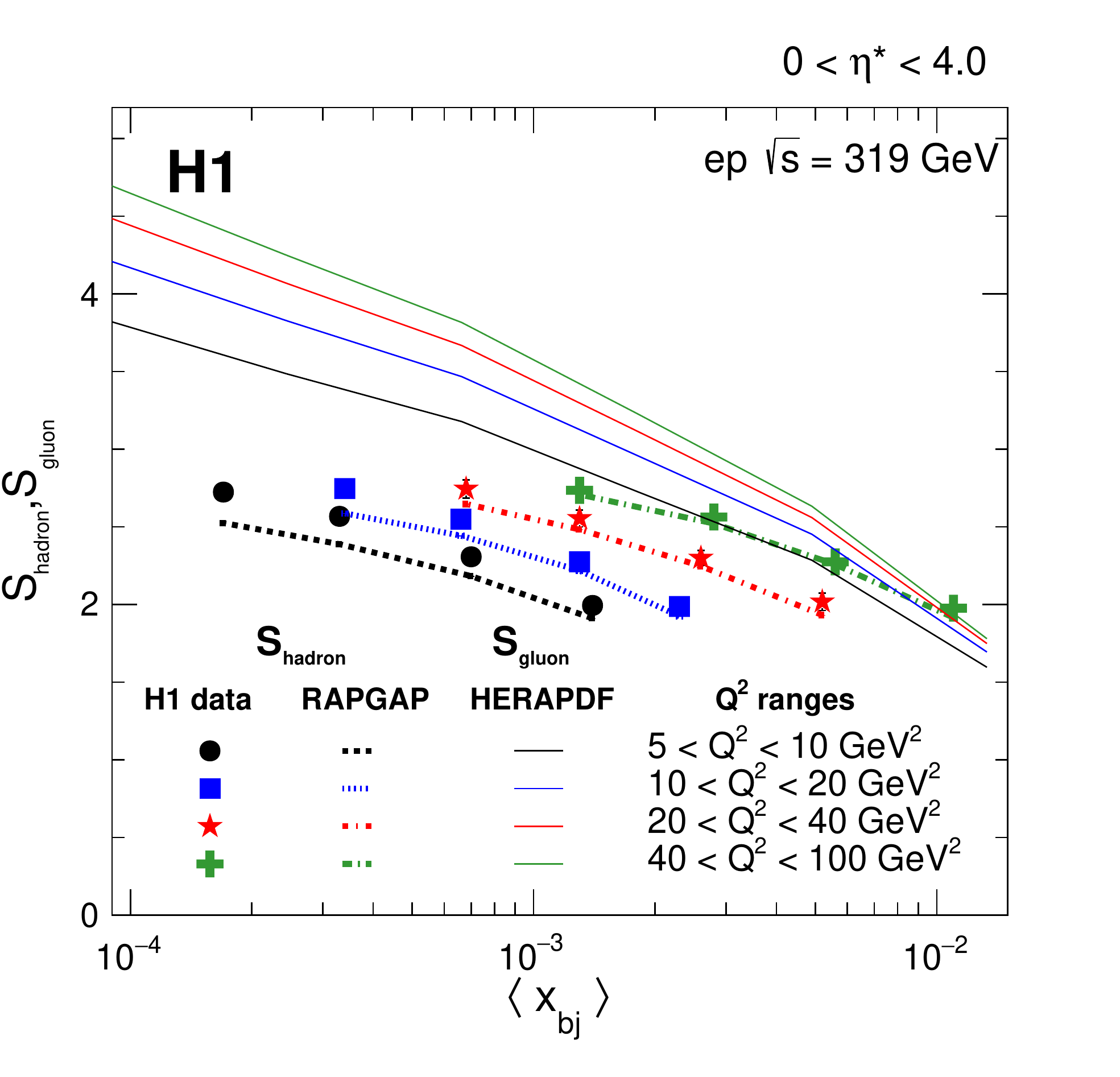
}
\caption{ \label{fig:figure_5_15} Hadron entropy $S_{\rm hadron}$ derived from multiplicity distributions as a function of $\langle$\xbj$\rangle$ measured in different \qsq ranges, measured in $\sqrt{s}=319$\gev~$ep$ collisions. Here, a restriction to the current hemisphere $0<\eta^{\ast}<4$ is applied. Further phase space restrictions are given in Table~\ref{tab:phasespace}.
  Predictions for $S_{\rm hadron}$ from the RAPGAP model and for the entanglement entropy $S_{\rm gluon}$ based on an entanglement model are shown by the dashed lines and solid lines, respectively. For each \qsq range, the value of the lower boundary is used for predicting $S_{\rm gluon}$.
  The total uncertainty on the data is represented by the error bars. }
\end{figure}

To investigate further, the hadron entropy determined in the current hemisphere,  $0<\eta^\ast<4$, is presented in figure~\ref{fig:figure_5_15} and Table~\ref{tab:table_figure_11}. The hadron entropy based on multiplicity distributions is shown as a function of $\langle$\xbj$\rangle$ in different bins of \qsqnospace. In contrast to Figure~\ref{fig:figure_5_14}, there is no moving pseudorapidity range applied. Independent of \qsqnospace, the measured hadron entropy, $S_{\rm hadron}$, rises with decreasing $\langle$\xbj$\rangle$ with similar slopes in different \qsq bins. Also shown are the predictions from RAPGAP which closely follow the data at high \qsq but underestimate the data at low \qsqnospace. Since $S_{\rm hadron}$ is calculated from the multiplicity distributions, this behaviour of the MC is related to what has been observed for the moments, figure~\ref{fig:figure_5_6} and~\ref{fig:figure_5_8}. The predictions from the entanglement approach based on the gluon density again fail to describe $S_{\rm hadron}$ in magnitude. However, at low \qsq the slope of $S_{\rm gluon}$ has some similarities with that observed for $S_{\rm hadron}$, while it becomes steeper than observed with increasing \qsqnospace.

\section{\label{sec:summary} Summary}
The charged particle multiplicity distributions $P(N)$ are measured in deep inelastic scattering at $\sqrt{s}=319$\gev~using the H1 detector at HERA. The integrated luminosity used in this analysis is $136$~\pbarnt, recorded in the years 2006 and 2007 in positrons scattering off protons. The $P(N)$ distributions are measured in bins of \qsqnospace, $y$ and pseudorapidity $\eta$. Predictions based on simulations of partonic tree level matrix elements, supplemented by parton shower and hadronisation are generally found to be consistent with the measurement at low multiplicities while they underestimate the data in the high multiplicity regions, especially at low \qsqnospace. For measurements of moments, the predictions generally describe the data well at low hadronic centre-of-mass energy $W$ and high \qsq but less so at high $W$ and low \qsqnospace. In addition, KNO scaling is observed within the kinematic phase space of the analysis. 

The measurement of the  charged particle multiplicity distributions is also used to test for the first time predictions based on quantum entanglement on sub-nucleonic scales in deep-inelastic $ep$ scattering. The predictions from the entropy of gluons are found to grossly disagree with the hadron entropy obtained from the multiplicity measurements presented here, and therefore the data do not support the basic concept of equality of the parton and hadron entropy with the current level of theory development. 

The measurements reported in this paper not only provide valuable information for better understanding particle production mechanisms, but also set an important testing ground for the development of new concepts, like quantum entanglement at sub-nucleonic scales.

\section*{Acknowledgements}
We are grateful to the HERA machine group whose outstanding efforts have made this experiment possible. We thank the engineers and technicians for their work in constructing and maintaining the H1 detector, our funding agencies for financial support, the DESY technical staff for continual assistance and the DESY directorate for support and for the hospitality which they extend to the non–DESY members of the collaboration.

We would like to give credit to all partners contributing to the EGI computing infrastructure for their support for the H1 Collaboration. 

We express our thanks to all those involved in securing not only the H1 data but also the software and working environment for long term use, allowing the unique H1 data set to continue to be explored in the coming years. The transfer from experiment specific to central resources with long term support, including both storage and batch systems, has also been crucial to this enterprise. We therefore also acknowledge the role played by DESY-IT and all people involved during this transition and their future role in the years to come.

\bibliography{desy20-176}

\providecommand{\href}[2]{#2}\begingroup\raggedright\begin{thebibliography}{10}

\bibitem{Bjorken:1969ja}
J.~D. Bjorken and E.~A. Paschos
\href{http://dx.doi.org/10.1103/PhysRev.185.1975}{{\em Phys. Rev.} {185} (1969)
  1975}.

\bibitem{Feynman:1969wa}
R.~P. Feynman
{\em Conf. Proc.} {C690905} (1969) 237.

\bibitem{Gribov:1968fc}
V.~N. Gribov
{\em Sov. Phys. JETP} {26} (1968) 414.

\bibitem{Klebanov:2007ws}
I.~R. Klebanov, D.~Kutasov and A.~Murugan
  \href{http://dx.doi.org/10.1016/j.nuclphysb.2007.12.017}{{\em Nucl. Phys.}
  {B796} (2008) 274},
\href{http://arxiv.org/abs/0709.2140}{{\ttfamily [arXiv:0709.2140]}}.

\bibitem{Kharzeev:2017qzs}
D.~E. Kharzeev and E.~M. Levin
  \href{http://dx.doi.org/10.1103/PhysRevD.95.114008}{{\em Phys. Rev.} {D95}
  (2017) 114008},
\href{http://arxiv.org/abs/1702.03489}{{\ttfamily [arXiv:1702.03489]}}.

\bibitem{Duan:2020jkz}
H.~Duan, C.~Akkaya, A.~Kovner and V.~V. Skokov
\href{http://arxiv.org/abs/2001.01726}{{\ttfamily [arXiv:2001.01726]}}.

\bibitem{Armesto:2019mna}
N.~Armesto, F.~Dominguez, A.~Kovner, M.~Lublinsky and V.~Skokov
  \href{http://dx.doi.org/10.1007/JHEP05(2019)025}{{\em JHEP} {05} (2019) 025},
\href{http://arxiv.org/abs/1901.08080}{{\ttfamily [arXiv:1901.08080]}}.

\bibitem{Kovner:2015hga}
A.~Kovner and M.~Lublinsky
  \href{http://dx.doi.org/10.1103/PhysRevD.92.034016}{{\em Phys. Rev.} {D92}
  (2015) 034016},
\href{http://arxiv.org/abs/1506.05394}{{\ttfamily [arXiv:1506.05394]}}.

\bibitem{Baker:2017wtt}
O.~K. Baker and D.~E. Kharzeev
  \href{http://dx.doi.org/10.1103/PhysRevD.98.054007}{{\em Phys. Rev.} {D98}
  (2018) 054007},
\href{http://arxiv.org/abs/1712.04558}{{\ttfamily [arXiv:1712.04558]}}.

\bibitem{Tu:2019ouv}
Z.~Tu, D.~E. Kharzeev and T.~Ullrich
  \href{http://dx.doi.org/10.1103/PhysRevLett.124.062001}{{\em Phys.\ Rev.\
  Lett.} {124} (2020) 062001},
  \href{http://arxiv.org/abs/1904.11974}{{\ttfamily [arXiv:1904.11974]}}.

\bibitem{Derrick:1995ca}
{ ZEUS} Collaboration, M.~Derrick {\em et~al.}
  \href{http://dx.doi.org/10.1007/BF01564824}{{\em Z. Phys. C} {67} (1995) 93},
  \href{http://arxiv.org/abs/hep-ex/9501012}{{\ttfamily [hep-ex/9501012]}}.

\bibitem{Aid:1996cb}
{ H1} Collaboration, S.~Aid {\em et~al.}
  \href{http://dx.doi.org/10.1007/s002880050280, 10.1007/BF02909189}{{\em Z.
  Phys.} {C72} (1996) 573},
\href{http://arxiv.org/abs/hep-ex/9608011}{{\ttfamily [hep-ex/9608011]}}.

\bibitem{Adloff:1997fr}
{ H1} Collaboration, C.~Adloff {\em et~al.}
  \href{http://dx.doi.org/10.1016/S0550-3213(97)00585-3}{{\em Nucl. Phys. B}
  {504} (1997) 3},
\href{http://arxiv.org/abs/hep-ex/9707005}{{\ttfamily [hep-ex/9707005]}}.

\bibitem{Adloff:1998dw}
{ H1} Collaboration, C.~Adloff {\em et~al.}
  \href{http://dx.doi.org/10.1007/s100529800963}{{\em Eur. Phys. J. C} {5}
  (1998) 439},
\href{http://arxiv.org/abs/hep-ex/9804012}{{\ttfamily [hep-ex/9804012]}}.

\bibitem{Chekanov:2008ae}
{ ZEUS} Collaboration, S.~Chekanov {\em et~al.}
  \href{http://dx.doi.org/10.1088/1126-6708/2008/06/061}{{\em JHEP} {06} (2008)
  061},
\href{http://arxiv.org/abs/0803.3878}{{\ttfamily [arXiv:0803.3878]}}.

\bibitem{Koba:1972ng}
Z.~Koba, H.~B. Nielsen and P.~Olesen
\href{http://dx.doi.org/10.1016/0550-3213(72)90551-2}{{\em Nucl. Phys.} {B40}
  (1972) 317}.

\bibitem{Mueller:1999fp}
A.~H. Mueller \href{http://dx.doi.org/10.1016/S0550-3213(99)00502-7}{{\em Nucl.
  Phys.} {B572} (2000) 227},
\href{http://arxiv.org/abs/hep-ph/9906322}{{\ttfamily [hep-ph/9906322]}}.

\bibitem{Dokshitzer_1991}
Y.~L. Dokshitzer, V.~A. Khoze and S.~I. Troyan
  \href{http://dx.doi.org/10.1088/0954-3899/17/10/017}{{\em Journal of Physics
  G: Nuclear and Particle Physics} {17} (1991) 1585}.

\bibitem{Jung:1993gf}
H.~Jung
\href{http://dx.doi.org/10.1016/0010-4655(94)00150-Z}{{\em Comput. Phys.
  Commun.} {86} (1995) 147}.

\bibitem{Charchula:1994kf}
K.~Charchula, G.~A. Schuler and H.~Spiesberger
\href{http://dx.doi.org/10.1016/0010-4655(94)90086-8}{{\em Comput. Phys.
  Commun.} {81} (1994) 381}.

\bibitem{Sjostrand:2014zea}
T.~Sj\"ostrand, S.~Ask, J.~R. Christiansen, R.~Corke, N.~Desai, P.~Ilten,
  S.~Mrenna, S.~Prestel, C.~O. Rasmussen and P.~Z. Skands
  \href{http://dx.doi.org/10.1016/j.cpc.2015.01.024}{{\em Comput. Phys.
  Commun.} {191} (2015) 159},
\href{http://arxiv.org/abs/1410.3012}{{\ttfamily [arXiv:1410.3012]}}.

\bibitem{Abt:1996hi}
{ H1} Collaboration, I.~Abt {\em et~al.}
\href{http://dx.doi.org/10.1016/S0168-9002(96)00893-5}{{\em Nucl. Instrum.
  Meth.} {A386} (1997) 310}.

\bibitem{Abt:1996xv}
{ H1} Collaboration, I.~Abt {\em et~al.}
\href{http://dx.doi.org/10.1016/S0168-9002(96)00894-7}{{\em Nucl. Instrum.
  Meth.} {A386} (1997) 348}.

\bibitem{Appuhn:1996na}
{ H1 SPACAL Group}, R.~D. Appuhn {\em et~al.}
\href{http://dx.doi.org/10.1016/S0168-9002(96)01171-0}{{\em Nucl. Instrum.
  Meth.} {A386} (1997) 397}.

\bibitem{Pitzl:2000wz}
D.~Pitzl {\em et~al.}
  \href{http://dx.doi.org/10.1016/S0168-9002(00)00488-5}{{\em Nucl. Instrum.
  Meth.} {A454} (2000) 334},
\href{http://arxiv.org/abs/hep-ex/0002044}{{\ttfamily [hep-ex/0002044]}}.

\bibitem{List:2002bd}
B.~List
\href{http://dx.doi.org/10.1016/S0168-9002(02)02009-0}{{\em Nucl. Instrum.
  Meth.} {A501} (2001) 49}.

\bibitem{Becker:2007ms}
J.~Becker {\em et~al.} \href{http://dx.doi.org/10.1016/j.nima.2007.11.024}{{\em
  Nucl. Instrum. Meth.} {A586} (2008) 190},
\href{http://arxiv.org/abs/physics/0701002}{{\ttfamily [physics/0701002]}}.

\bibitem{Laycock:2012xg}
P.~J. Laycock, R.~C.~W. Henderson, S.~J. Maxfield, J.~V. Morris, G.~D. Patel
  and D.~P.~C. Sankey
  \href{http://dx.doi.org/10.1088/1748-0221/7/08/T08003}{{\em JINST} {7} (2012)
  T08003},
\href{http://arxiv.org/abs/1206.4068}{{\ttfamily [arXiv:1206.4068]}}.

\bibitem{Nicholls:1995di}
{ H1 SPACAL Group}, T.~Nicholls {\em et~al.}
\href{http://dx.doi.org/10.1016/0168-9002(95)01443-8}{{\em Nucl. Instrum.
  Meth.} {A374} (1996) 149}.

\bibitem{Andrieu:1993kh}
{ H1 Calorimeter Group}, B.~Andrieu {\em et~al.}
\href{http://dx.doi.org/10.1016/0168-9002(93)91257-N}{{\em Nucl. Instrum.
  Meth.} {A336} (1993) 460}.

\bibitem{Andrieu:1993tz}
{ H1 Calorimeter Group}, B.~Andrieu {\em et~al.}
\href{http://dx.doi.org/10.1016/0168-9002(93)91258-O}{{\em Nucl. Instrum.
  Meth.} {A336} (1993) 499}.

\bibitem{Gribov:1972ri}
V.~N. Gribov and L.~N. Lipatov
{\em Sov. J. Nucl. Phys.} {15} (1972) 438.

\bibitem{Lipatov:1974qm}
L.~N. Lipatov
{\em Sov. J. Nucl. Phys.} {20} (1975) 94.

\bibitem{Altarelli:1977zs}
G.~Altarelli and G.~Parisi
\href{http://dx.doi.org/10.1016/0550-3213(77)90384-4}{{\em Nucl. Phys.} {B126}
  (1977) 298}.

\bibitem{Dokshitzer:1977sg}
Y.~L. Dokshitzer
{\em Sov. Phys. JETP} {46} (1977) 641.

\bibitem{Pumplin:2002vw}
J.~Pumplin, D.~Stump, J.~Huston, H.~Lai, P.~M. Nadolsky and W.~Tung
  \href{http://dx.doi.org/10.1088/1126-6708/2002/07/012}{{\em JHEP} {07} (2002)
  012}, \href{http://arxiv.org/abs/hep-ph/0201195}{{\ttfamily
  [hep-ph/0201195]}}.

\bibitem{Lonnblad:1992tz}
L.~L\"onnblad
\href{http://dx.doi.org/10.1016/0010-4655(92)90068-A}{{\em Comput. Phys.
  Commun.} {71} (1992) 15}.

\bibitem{Kuraev:1976ge}
E.~A. Kuraev, L.~N. Lipatov and V.~S. Fadin
{\em Sov. Phys. JETP} {44} (1976) 443.

\bibitem{Kuraev:1977fs}
E.~A. Kuraev, L.~N. Lipatov and V.~S. Fadin
{\em Sov. Phys. JETP} {45} (1977) 199.

\bibitem{Balitsky:1978ic}
I.~I. Balitsky and L.~N. Lipatov
{\em Sov. J. Nucl. Phys.} {28} (1978) 822.

\bibitem{Lonnblad:1994wk}
L.~L\"onnblad
\href{http://dx.doi.org/10.1007/BF01571885}{{\em Z. Phys.} {C65} (1995) 285}.

\bibitem{Hoche:2015sya}
S.~Höche and S.~Prestel
  \href{http://dx.doi.org/10.1140/epjc/s10052-015-3684-2}{{\em Eur. Phys. J.}
  {C75} (2015) 461},
\href{http://arxiv.org/abs/1506.05057}{{\ttfamily [arXiv:1506.05057]}}.

\bibitem{Sjostrand:2001yu}
T.~Sj\"ostrand, L.~L\"onnblad and S.~Mrenna
\href{http://arxiv.org/abs/hep-ph/0108264}{{\ttfamily [hep-ph/0108264]}}.

\bibitem{Spiesberger:1992vu}
H.~Spiesberger {\em et~al.}, ``{Radiative corrections at HERA},'' in {\em
  {Workshop on Physics at HERA}}, 3 1992.

\bibitem{Kwiatkowski:1990es}
A.~Kwiatkowski, H.~Spiesberger and H.~J. Mohring
\href{http://dx.doi.org/10.1016/0010-4655(92)90136-M}{{\em Comput. Phys.
  Commun.} {69} (1992) 155}.

\bibitem{Kwiatkowski:1990cx}
A.~Kwiatkowski, H.~Spiesberger and H.~Mohring
  \href{http://dx.doi.org/10.1007/BF01558572}{{\em Z. Phys. C} {50} (1991)
  165}.

\bibitem{Brun:1994aa}
R.~Brun, F.~Bruyant, F.~Carminati, S.~Giani, M.~Maire, A.~McPherson, G.~Patrick
  and L.~Urban.

\bibitem{Aaron:2012kn}
{ H1} Collaboration, F.~D. Aaron {\em et~al.}
  \href{http://dx.doi.org/10.1140/epjc/s10052-014-2733-6}{{\em Eur. Phys. J.}
  {C72} (2012) 2163}, \href{http://arxiv.org/abs/1205.2448}{{\ttfamily
  [arXiv:1205.2448]}}.
[Erratum: Eur. Phys. J. C74 (2012) 2733].

\bibitem{Peez:2003zd}
M.~Peez, \href{http://dx.doi.org/10.3204/DESY-THESIS-2003-023}{{\em Search for
  deviations from the standard model in high transverse energy processes at the
  electron proton collider HERA}}.
\newblock Thesis, Lyon U., (2003) .

\bibitem{Hellwig:2004bb}
S.~Hellwig, {\em {Untersuchung der D$\ast - \pi$ slow Double Tagging Methode in
  Charmanalysen}}.
\newblock Thesis, Hamburg U., 2005.
\newblock \url{http://www-h1.desy.de/publications/theses_list.html}.

\bibitem{Portheault:2005uu}
B.~Portheault, {\em {First measurement of charged and neutral current cross
  sections with the polarized positron beam at HERA II and QCD-electroweak
  analyses}}.
\newblock Thesis, Paris XI U., 2004.
\newblock \url{http://www-h1.desy.de/publications/theses_list.html}.

\bibitem{Alexa:2013vkv}
{ H1} Collaboration, C.~Alexa {\em et~al.}
  \href{http://dx.doi.org/10.1140/epjc/s10052-013-2406-x}{{\em Eur. Phys. J.}
  {C73} (2013) 2406},
\href{http://arxiv.org/abs/1302.1321}{{\ttfamily [arXiv:1302.1321]}}.

\bibitem{Bassler:1994uq}
U.~Bassler and G.~Bernardi
  \href{http://dx.doi.org/10.1016/0168-9002(95)00173-5}{{\em Nucl. Instrum.
  Meth. A} {361} (1995) 197},
  \href{http://arxiv.org/abs/hep-ex/9412004}{{\ttfamily [hep-ex/9412004]}}.

\bibitem{Sjostrand:2006za}
T.~Sj\"ostrand, S.~Mrenna and P.~Skands
  \href{http://dx.doi.org/10.1088/1126-6708/2006/05/026}{{\em JHEP} {05} (2006)
  026},
\href{http://arxiv.org/abs/hep-ph/0603175}{{\ttfamily [hep-ph/0603175]}}.

\bibitem{Schmitt:2012kp}
S.~Schmitt \href{http://dx.doi.org/10.1088/1748-0221/7/10/T10003}{{\em JINST}
  {7} (2012) T10003},
\href{http://arxiv.org/abs/1205.6201}{{\ttfamily [arXiv:1205.6201]}}.

\bibitem{Aaron:2009bp}
{ H1} Collaboration, F.~Aaron {\em et~al.}
  \href{http://dx.doi.org/10.1140/epjc/s10052-009-1128-6}{{\em Eur. Phys. J. C}
  {63} (2009) 625}, \href{http://arxiv.org/abs/0904.0929}{{\ttfamily
  [arXiv:0904.0929]}}.

\bibitem{Salek:2010ifa}
D.~\v{S}\'alek, {\em {Measurement of the Longitudinal Proton Structure Function
  in Diffraction at the H1 Experiment and Prospects for Diffraction at LHC}}.
\newblock PhD thesis, Charles U., (2010) .
\newblock \url{http://www-h1.desy.de/publications/theses_list.html}.

\bibitem{Aad:2016xww}
{ ATLAS} Collaboration, G.~Aad {\em et~al.}
  \href{http://dx.doi.org/10.1140/epjc/s10052-016-4203-9}{{\em Eur. Phys. J.}
  {C76} (2016) 403},
\href{http://arxiv.org/abs/1603.02439}{{\ttfamily [arXiv:1603.02439]}}.

\bibitem{ALICE:2017pcy}
{ ALICE} Collaboration, S.~Acharya {\em et~al.}
  \href{http://dx.doi.org/10.1140/epjc/s10052-017-5412-6}{{\em Eur. Phys. J.}
  {C77} (2017) 852},
\href{http://arxiv.org/abs/1708.01435}{{\ttfamily [arXiv:1708.01435]}}.

\bibitem{Khachatryan:2010nk}
{ CMS} Collaboration, V.~Khachatryan {\em et~al.}
  \href{http://dx.doi.org/10.1007/JHEP01(2011)079}{{\em JHEP} {01} (2011) 079},
\href{http://arxiv.org/abs/1011.5531}{{\ttfamily [arXiv:1011.5531]}}.

\bibitem{Aaij:2014pza}
{ LHCb} Collaboration, R.~Aaij {\em et~al.}
  \href{http://dx.doi.org/10.1140/epjc/s10052-014-2888-1}{{\em Eur. Phys. J.}
  {C74} (2014) 2888},
\href{http://arxiv.org/abs/1402.4430}{{\ttfamily [arXiv:1402.4430]}}.

\bibitem{Abramowicz:2015mha}
{ H1 and ZEUS} Collaboration, H.~Abramowicz {\em et~al.}
  \href{http://dx.doi.org/10.1140/epjc/s10052-015-3710-4}{{\em Eur. Phys. J.}
  {C75} (2015) 580},
\href{http://arxiv.org/abs/1506.06042}{{\ttfamily [arXiv:1506.06042]}}.

\end{thebibliography}\endgroup
\bibliographystyle{desy20-176}

\clearpage

\begin{figure}[tbh]
  \centering
\includegraphics[width=5.8in]{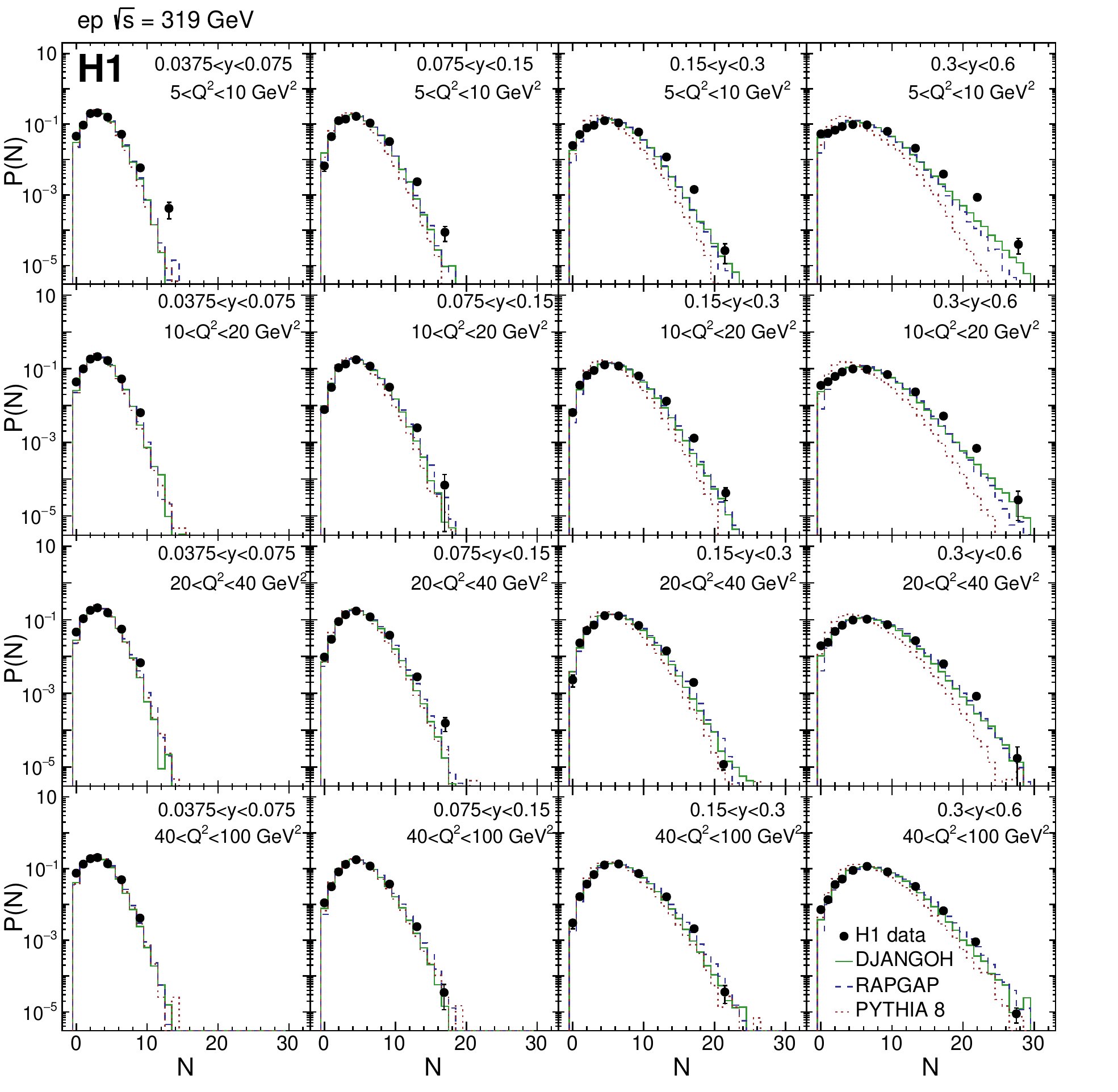
}
  \caption{ \label{fig:figure_5_1} Charged particle multiplicity distributions $P(N)$ as a function of the number of particles $N$ at $\sqrt{s}=319$\gev~ep collisions. Different panels correspond to different \qsq and $y$ bins, as indicated by the text in the figure. The phase space restrictions are given in Table~\ref{tab:phasespace}. Predictions from DJANGOH, RAPGAP and PYTHIA 8 are also shown. The total uncertainty is denoted by the error bars.}
\end{figure}

\begin{figure}[tbh]
  \centering
\includegraphics[width=5.8in]{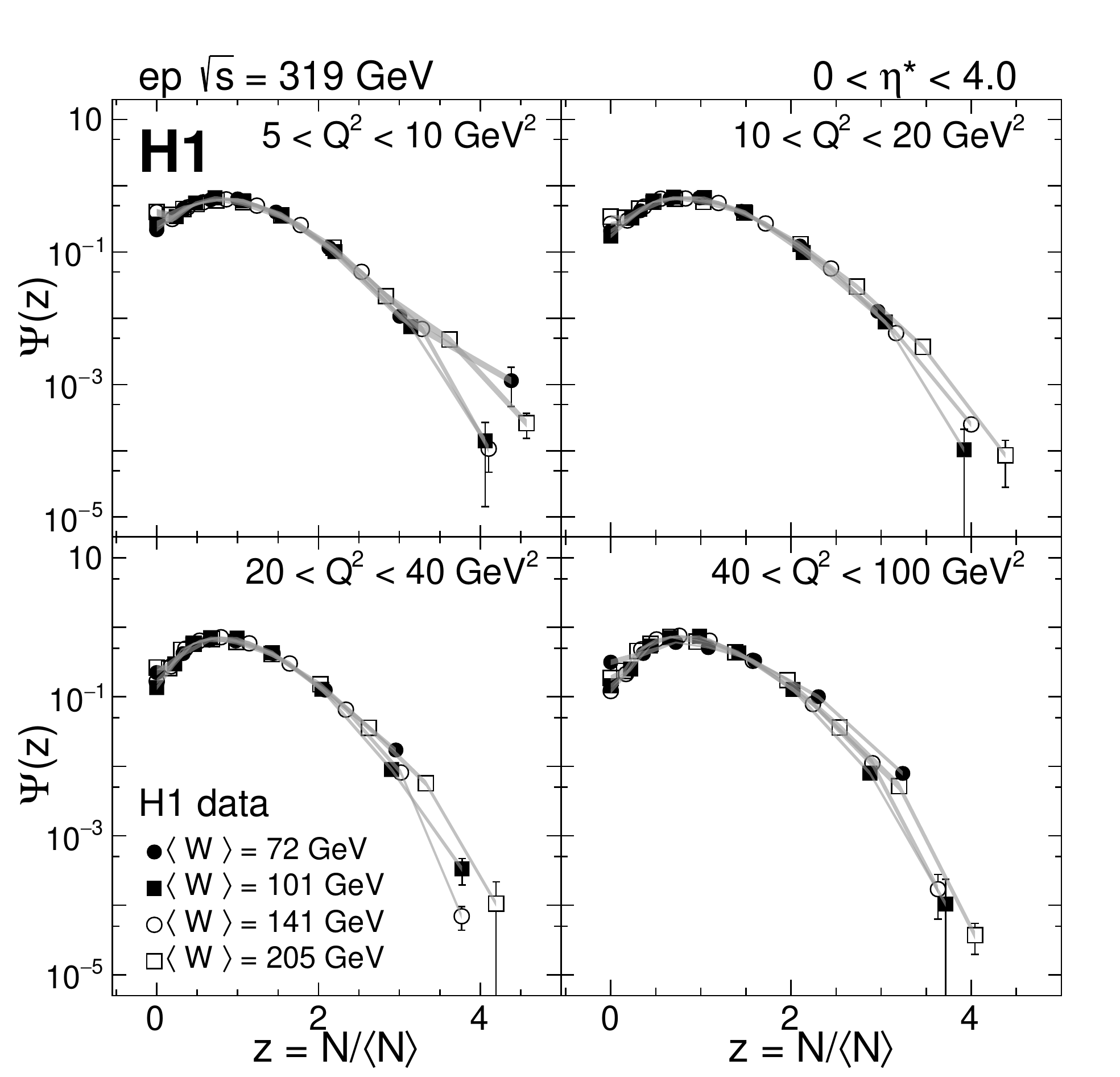
}
\caption{ \label{fig:figure_5_10_KNO} KNO function $\Psi(z)$ as a function of $z$ measured at $\sqrt{s}=319$\gev~in $ep$ collisions in bins of \qsq with an additional restriction to the current hemisphere $0<\eta^{\ast}<4$. Further phase space restrictions are given in Table~\ref{tab:phasespace}. The total uncertainty is denoted by the error bars.}
\end{figure}

 \begin{table}[tb] 
 \fontsize{8}{8}\selectfont
  \begin{center} 
\begin{tabular}{|c|rrrrrrrrrrr|} 
   \multicolumn{12}{c}{\bf \normalsize \boldmath Mean multiplicity in DIS} \\
   \multicolumn{12}{c}{ \normalsize $|\eta_{_{\rm{lab}}}|<1.6$ } \\
   \noalign{\vskip 2mm} 
   \hline 
   \multicolumn{12}{|c|}{ \boldmath $5<\qsq<10~\gevsq$ } \\
   \hline 
    & & stat. & sys.tot & sys.1 & sys.2 & sys.3 & sys.4 & sys.5 & sys.6 & sys.7  & sys.8  \\ $\langle W \rangle$ & $\langle N \rangle$ & (\%) & (\%) & (\%) & (\%) & (\%) & (\%) & (\%) & (\%) & (\%) & (\%)  \\
   \hline 
72      & 3.4 & 0.39 & 4.7 & 1 & 3 & 2 & 2 & 1 & 1.5 & 0.5 & 1  \\
101      & 4.8 & 0.2 & 3.8 & 1 & 1 & 2 & 2 & 1 & 1.5 & 0.5 & 1  \\
147      & 5.9 & 0.15 & 3.8 & 1 & 1 & 2 & 2 & 1 & 1.5 & 0.5 & 1  \\
\hline 
 \end{tabular} 
 \caption{Mean multiplicity measured in $ep$ DIS at $\sqrt{s}=319$\gev as a function of $W$. The measurement is repeated in four ranges of \qsq as indicated. The phase-space is further restricted as shown in table \ref{tab:phasespace}.}
 \label{tab:table_figure_8_a} 
  \end{center} 
   \end{table} 

 \begin{table}[tb] 
 \fontsize{8}{8}\selectfont
  \begin{center} 
   \begin{tabular}{|c|rrrrrrrrrrr|} 
   \multicolumn{12}{c}{\bf \normalsize \boldmath Mean multiplicity in DIS} \\
   \multicolumn{12}{c}{ \normalsize $0<\rap<4$ } \\
   \noalign{\vskip 2mm} 
   \hline 
   \multicolumn{12}{|c|}{ \boldmath $5<\qsq<10~\gevsq$ } \\
   \hline 
    & & stat. & sys.tot & sys.1 & sys.2 & sys.3 & sys.4 & sys.5 & sys.6 & sys.7  & sys.8  \\ $\langle W \rangle$ & $\langle N \rangle$ & (\%) & (\%) & (\%) & (\%) & (\%) & (\%) & (\%) & (\%) & (\%) & (\%)  \\
   \hline 
72      & 3 & 0.44 & 4.7 & 1 & 3 & 2 & 2 & 1 & 1.5 & 0.5 & 1  \\
101      & 4.2 & 0.21 & 3.8 & 1 & 1 & 2 & 2 & 1 & 1.5 & 0.5 & 1  \\
147      & 5.2 & 0.17 & 3.8 & 1 & 1 & 2 & 2 & 1 & 1.5 & 0.5 & 1  \\
\hline 
 \end{tabular} 
   \caption{Mean multiplicity measured in $ep$ DIS at $\sqrt{s}=319$\gev~as a function of $W$. The measurement is repeated in four ranges of \qsq as indicated. The phase-space is further restricted as shown in table \ref{tab:phasespace}.
    For this dataset, the track pseudorapidities in the hadronic centre-of-mass frame are restricted to the range $0<\rap<4$.} 
 \label{tab:table_figure_8_b} 
  \end{center} 
   \end{table} 

 \begin{table}[tb] 
 \fontsize{8}{8}\selectfont
  \begin{center} 
   \begin{tabular}{|c|rrrrrrrrrrr|} 
   \multicolumn{12}{c}{\bf \normalsize \boldmath Variance of multiplicity distributions in DIS} \\
   \multicolumn{12}{c}{ \normalsize $|\eta_{_{\rm{lab}}}|<1.6$ } \\
   \noalign{\vskip 2mm} 
   \hline 
   \multicolumn{12}{|c|}{ \boldmath $5<\qsq<10~\gevsq$ } \\
   \hline 
    & & stat. & sys.tot & sys.1 & sys.2 & sys.3 & sys.4 & sys.5 & sys.6 & sys.7  & sys.8  \\ $\langle W \rangle$ & $Var(N)$ & (\%) & (\%) & (\%) & (\%) & (\%) & (\%) & (\%) & (\%) & (\%) & (\%)  \\
   \hline 
72      & 3.6 & 2.3 & 6.4 & 1 & 5 & 2 & 2 & 1 & 2 & 1 & 1  \\
101      & 5.8 & 0.71 & 4.5 & 1 & 2 & 2 & 2 & 1 & 2 & 1 & 1  \\
147      & 10 & 0.42 & 6.4 & 1 & 5 & 2 & 2 & 1 & 2 & 1 & 1  \\
\hline 
 \end{tabular} 
   \caption{Variance of multiplicity distributions measured in $ep$ DIS at $\sqrt{s}=319$\gev~as a function of $W$. The measurement is repeated in four ranges of \qsq as indicated. The phase-space is further restricted as shown in table \ref{tab:phasespace}.}
 \label{tab:table_figure_9_a} 
  \end{center} 
   \end{table} 

 \begin{table}[tb] 
 \fontsize{8}{8}\selectfont
  \begin{center} 
   \begin{tabular}{|c|rrrrrrrrrrr|} 
   \multicolumn{12}{c}{\bf \normalsize \boldmath Variance of multiplicity distributions in DIS} \\
   \multicolumn{12}{c}{ \normalsize $0<\rap<4$ } \\
   \noalign{\vskip 2mm} 
   \hline 
   \multicolumn{12}{|c|}{ \boldmath $5<\qsq<10~\gevsq$ } \\
   \hline 
    & & stat. & sys.tot & sys.1 & sys.2 & sys.3 & sys.4 & sys.5 & sys.6 & sys.7  & sys.8  \\ $\langle W \rangle$ & $Var(N)$ & (\%) & (\%) & (\%) & (\%) & (\%) & (\%) & (\%) & (\%) & (\%) & (\%)  \\
   \hline 
72      & 3.5 & 2.8 & 6.4 & 1 & 5 & 2 & 2 & 1 & 2 & 1 & 1  \\
101      & 6.6 & 0.53 & 4.5 & 1 & 2 & 2 & 2 & 1 & 2 & 1 & 1  \\
147      & 11 & 0.41 & 6.4 & 1 & 5 & 2 & 2 & 1 & 2 & 1 & 1  \\
\hline 
 \end{tabular} 
 \caption{Variance of multiplicity distributions measured in $ep$ DIS at $\sqrt{s}=319$\gev~as a function of $W$. The measurement is repeated in four ranges of \qsq as indicated. The phase-space is further restricted as shown in table \ref{tab:phasespace}.
     For this dataset, the track pseudorapidities in the hadronic centre-of-mass frame are restricted to the range $0<\rap<4$.}  
 \label{tab:table_figure_9_b} 
  \end{center} 
   \end{table} 

 \begin{table}[tb] 
 \fontsize{7}{8}\selectfont
  \begin{center} 
   \begin{tabular}{|c|c|rrrrrrrrrrr|} 
   \multicolumn{13}{c}{\bf \normalsize \boldmath Hadron entropy of multiplicity distributions in DIS} \\
   \multicolumn{13}{c}{ } \\
   \noalign{\vskip 2mm} 
   \hline 
   \multicolumn{13}{|c|}{ \boldmath $5<\qsq<10~\gevsq$ } \\
   \hline 
   &  &  & stat. & sys.tot & sys.1 & sys.2 & sys.3 & sys.4 & sys.5 & sys.6 & sys.7 & sys.8  \\   $\langle x_{\rm bj} \rangle$ & $\eta_{\rm lab}$ & $S_{\rm{hadron}}$ & (\%) & (\%) & (\%) & (\%) & (\%) & (\%) & (\%) & (\%) & (\%) & (\%)  \\
   \hline 
0.00033 & (-1.2,0.2)      & 2 & 0.19 & 2.7 & 0 & 2 & 1 & 1 & 0.5 & 0.5 & 0 & 1  \\
0.0007 & (-0.5,0.9)      & 1.9 & 0.2 & 2.1 & 0 & 1 & 1 & 1 & 0.5 & 0.5 & 0 & 1  \\
0.0014 & (0.2,1.6)      & 1.9 & 0.33 & 2.1 & 0 & 1 & 1 & 1 & 0.5 & 0.5 & 0 & 1  \\
\hline 
 \end{tabular} 
 \caption{Hadron entropy derived from multiplicity distributions measured in $ep$ DIS at $\sqrt{s}=319$\gev~as a function of $\langle x_{\rm bj} \rangle$. The measurement is repeated in four ranges of \qsq as indicated. The phase-space is further restricted as shown in table \ref{tab:phasespace}. For this dataset, the entropy is determined in a sliding window in pseudorapidity as indicated in the table.}
 \label{tab:table_figure_10} 
  \end{center} 
   \end{table} 

 \begin{table}[tb] 
 \fontsize{8}{8}\selectfont
  \begin{center} 
   \begin{tabular}{|c|rrrrrrrrrrr|} 
   \multicolumn{12}{c}{\bf \normalsize \boldmath Hadron entropy of multiplicity distributions in DIS} \\
   \multicolumn{12}{c}{ \normalsize $0<\rap<4$ } \\
   \noalign{\vskip 2mm} 
   \hline 
   \multicolumn{12}{|c|}{ \boldmath $5<\qsq<10~\gevsq$ } \\
   \hline 
    &  & stat. & sys.tot & sys.1 & sys.2 & sys.3 & sys.4 & sys.5 & sys.6 & sys.7 & sys.8  \\ $\langle x_{\rm bj} \rangle$ & $S_{\rm{hadron}}$ & (\%) & (\%) & (\%) & (\%) & (\%) & (\%) & (\%) & (\%) & (\%) & (\%)  \\
   \hline 
0.0014      & 2 & 0.33 & 2.7 & 0 & 2 & 1 & 1 & 0.5 & 0.5 & 0 & 1  \\
0.0007      & 2.3 & 0.089 & 2.1 & 0 & 1 & 1 & 1 & 0.5 & 0.5 & 0 & 1  \\
0.00033      & 2.6 & 0.06 & 2.1 & 0 & 1 & 1 & 1 & 0.5 & 0.5 & 0 & 1  \\
\hline 
 \end{tabular} 
 \caption{Hadron entropy derived from multiplicity distributions measured in $ep$ DIS at $\sqrt{s}=319$\gev~as a function of $\langle x_{\rm bj} \rangle$.  The measurement is repeated in four ranges of \qsq as indicated. The phase-space is further restricted as shown in table \ref{tab:phasespace}. For this dataset, the entropy is determined with the track pseudorapidities in the hadronic centre-of-mass frame are restricted to the range $0<\rap<4$} 
 \label{tab:table_figure_11} 
  \end{center} 
   \end{table} 

 \FloatBarrier
 \newpage
\appendix
\section{Charged particle multiplicities and KNO function}

Numerical values of the charged particle multiplicity measured in the pseudorapidity range $-1.6<\eta_{_{\text{lab}}}<1.6$ are presented in four ranges of \qsq and are further subdivided into four $y$ ranges. The four \qsq  ranges are shown in the tables \ref{tab:table_figure_1_a}, \ref{tab:table_figure_1_b}, \ref{tab:table_figure_1_c}, and \ref{tab:table_figure_1_d}, respectively. Further phase space restrictions are discussed in table \ref{tab:phasespace}. These restrictions apply to all measurements presented in the following.

The measurement is repeated in three overlapping subranges of 
$\eta_{_{\text{lab}}}$. There are a total of twelve tables, each corresponding to selected range in both \qsq and $\eta_{_{\text{lab}}}$. The range $5<\qsq<10~\gevsq$ is given in tables \ref{tab:table_figure_2_a}, \ref{tab:table_figure_2_b}, and \ref{tab:table_figure_2_c}. The range $10<\qsq<20~\gevsq$ is given in tables \ref{tab:table_figure_3_a}, \ref{tab:table_figure_3_b}, and \ref{tab:table_figure_3_c}.
The range $20<\qsq<40~\gevsq$ is given in tables \ref{tab:table_figure_4_a}, \ref{tab:table_figure_4_b}, and \ref{tab:table_figure_4_c}.
The range $40<\qsq<100~\gevsq$ is given in tables \ref{tab:table_figure_5_a}, \ref{tab:table_figure_5_b}, and \ref{tab:table_figure_5_c}.

The measurement is also repeated over the full available pseudorapidity range $-1.6<\eta_{_{\text{lab}}}<1.6$ in the laboratory frame, however with the extra restriction of the track selection to the current hemisphere $0<\eta^\ast<4$. Again the data are given in four \qsq ranges, shown in the tables \ref{tab:table_figure_6_a}, \ref{tab:table_figure_6_b}, \ref{tab:table_figure_6_c}, and \ref{tab:table_figure_6_d}, respectively. For these data, the KNO function $\Psi(z)$ is also determined. It is reported in the tables \ref{tab:table_figure_7_a}, \ref{tab:table_figure_7_b}, \ref{tab:table_figure_7_c}, and \ref{tab:table_figure_7_d}.

 \begin{table}[tb] 
 \fontsize{7.5}{7.5}\selectfont
  \begin{center} 
\begin{tabular}{|cc|c|rrrrrrrrrr|} 
   \multicolumn{13}{c}{\bf \normalsize \boldmath Multiplicity distributions in DIS} \\
   \multicolumn{13}{c}{ \normalsize $5<\qsq<10~\gevsq$ and $|\eta_{_{\rm{lab}}}| < 1.6$ } \\
   \noalign{\vskip 2mm} 
   \hline 
   \multicolumn{13}{|c|}{ \boldmath $0.0375<y<0.075$ } \\
   \hline 
   \multicolumn{2}{|c|}{ } & & & stat. & sys.tot & sys.1 & sys.2 & sys.3 & sys.4 & sys.5 & sys.6 & sys.7 \\
   \multicolumn{2}{|c|}{multiplicity range} & $\langle N \rangle$ & $P(N)$ & (\%) & (\%) & (\%) & (\%) & (\%) & (\%) & (\%) & (\%) & (\%) \\
   \hline 
-0.5 & 0.5 & 0      & 0.046 & 5.9 & 23 & 1 & 22 & 5 & 3 & 1 & 1 & 5  \\
0.5 & 1.5 & 1      & 0.094 & 2.9 & 23 & 1 & 22 & 5 & 3 & 1 & 1 & 5  \\
1.5 & 2.5 & 2      & 0.2 & 1.6 & 3.6 & 1 & 2 & 1 & 1 & 1 & 2 & 1  \\
\hline 
 \end{tabular} 
 \caption{Charged particle multiplicity distributions $P(N)$ measured in $ep$ DIS at $\sqrt{s}=319$\gev~with in the ranges of photon virtuality $5<\qsq<10~\gevsq$ and charged particle pseudorapidity range $|\eta_{_{\rm{lab}}}|<1.6$, presented in four ranges of inelasticity $y$ as indicated. The phase-space is further restricted as shown in table \ref{tab:phasespace}.} 
 \label{tab:table_figure_1_a} 
  \end{center} 
   \end{table} 

 \begin{table}[tb] 
 \fontsize{7.5}{7.5}\selectfont
  \begin{center} 
   \begin{tabular}{|cc|c|rrrrrrrrrr|} 
   \multicolumn{13}{c}{\bf \normalsize \boldmath Multiplicity distributions in DIS} \\
   \multicolumn{13}{c}{ \normalsize $10<\qsq<20~\gevsq$ and $|\eta_{_{\rm{lab}}}| < 1.6$ } \\
   \noalign{\vskip 2mm} 
   \hline 
   \multicolumn{13}{|c|}{ \boldmath $0.0375<y<0.075$ } \\
   \hline 
   \multicolumn{2}{|c|}{ } & & & stat. & sys.tot & sys.1 & sys.2 & sys.3 & sys.4 & sys.5 & sys.6 & sys.7 \\ \multicolumn{2}{|c|}{multiplicity range} & $\langle N \rangle$ & $P(N)$ & (\%) & (\%) & (\%) & (\%) & (\%) & (\%) & (\%) & (\%) & (\%) \\
   \hline 
-0.5 & 0.5 & 0      & 0.044 & 4.1 & 12 & 1 & 2 & 5 & 3 & 1 & 1 & 10  \\
0.5 & 1.5 & 1      & 0.099 & 2 & 12 & 1 & 2 & 5 & 3 & 1 & 1 & 10  \\
1.5 & 2.5 & 2      & 0.18 & 1.3 & 3.6 & 1 & 2 & 1 & 1 & 1 & 2 & 1  \\
\hline 
 \end{tabular} 
 \caption{Charged particle multiplicity distributions $P(N)$ measured in $ep$ DIS at $\sqrt{s}=319$\gev~with in the ranges of photon virtuality $10<\qsq<20~\gevsq$ and charged particle pseudorapidity range $|\eta_{_{\rm{lab}}}|<1.6$, presented in four ranges of inelasticity $y$ as indicated. The phase-space is further restricted as shown in table \ref{tab:phasespace}.} 
 \label{tab:table_figure_1_b} 
  \end{center} 
   \end{table} 
 \begin{table}[tb] 
 \fontsize{7.5}{7.5}\selectfont
  \begin{center} 
   \begin{tabular}{|cc|c|rrrrrrrrrr|} 
   \multicolumn{13}{c}{\bf \normalsize \boldmath Multiplicity distributions in DIS} \\
   \multicolumn{13}{c}{ \normalsize $20<\qsq<40~\gevsq$ and $|\eta_{_{\rm{lab}}}| < 1.6$ } \\
   \noalign{\vskip 2mm} 
   \hline 
   \multicolumn{13}{|c|}{ \boldmath $0.0375<y<0.075$ } \\
   \hline 
   \multicolumn{2}{|c|}{ } & & & stat. & sys.tot & sys.1 & sys.2 & sys.3 & sys.4 & sys.5 & sys.6 & sys.7 \\ \multicolumn{2}{|c|}{multiplicity range} & $\langle N \rangle$ & $P(N)$ & (\%) & (\%) & (\%) & (\%) & (\%) & (\%) & (\%) & (\%) & (\%) \\
   \hline 
-0.5 & 0.5 & 0      & 0.046 & 4.4 & 9.5 & 1 & 2 & 5 & 3 & 1 & 1 & 7  \\
0.5 & 1.5 & 1      & 0.11 & 2.6 & 9.5 & 1 & 2 & 5 & 3 & 1 & 1 & 7  \\
1.5 & 2.5 & 2      & 0.18 & 1.8 & 3.6 & 1 & 2 & 1 & 1 & 1 & 2 & 1  \\
\hline 
 \end{tabular} 
 \caption{Charged particle multiplicity distributions $P(N)$ measured in $ep$ DIS at $\sqrt{s}=319$\gev~with in the ranges of photon virtuality $20<\qsq<40~\gevsq$ and charged particle pseudorapidity range $|\eta_{_{\rm{lab}}}|<1.6$, presented in four ranges of inelasticity $y$ as indicated. The phase-space is further restricted as shown in table \ref{tab:phasespace}.} 
 \label{tab:table_figure_1_c} 
  \end{center} 
   \end{table} 
 \begin{table}[tb] 
 \fontsize{7.5}{7.5}\selectfont
  \begin{center} 
   \begin{tabular}{|cc|c|rrrrrrrrrr|} 
   \multicolumn{13}{c}{\bf \normalsize \boldmath Multiplicity distributions in DIS} \\
   \multicolumn{13}{c}{ \normalsize $40<\qsq<100~\gevsq$ and $|\eta_{_{\rm{lab}}}| < 1.6$ } \\
   \noalign{\vskip 2mm} 
   \hline 
   \multicolumn{13}{|c|}{ \boldmath $0.0375<y<0.075$ } \\
   \hline 
   \multicolumn{2}{|c|}{ } & & & stat. & sys.tot & sys.1 & sys.2 & sys.3 & sys.4 & sys.5 & sys.6 & sys.7 \\ \multicolumn{2}{|c|}{multiplicity range} & $\langle N \rangle$ & $P(N)$ & (\%) & (\%) & (\%) & (\%) & (\%) & (\%) & (\%) & (\%) & (\%) \\
   \hline 
-0.5 & 0.5 & 0      & 0.075 & 3.4 & 9.5 & 1 & 2 & 5 & 3 & 1 & 1 & 7  \\
0.5 & 1.5 & 1      & 0.13 & 2.8 & 9.5 & 1 & 2 & 5 & 3 & 1 & 1 & 7  \\
1.5 & 2.5 & 2      & 0.19 & 2.3 & 3.6 & 1 & 2 & 1 & 1 & 1 & 2 & 1  \\
\hline 
 \end{tabular} 
 \caption{Charged particle multiplicity distributions $P(N)$ measured in $ep$ DIS at $\sqrt{s}=319$\gev~with in the ranges of photon virtuality $40<\qsq<100~\gevsq$ and charged particle pseudorapidity range $|\eta_{_{\rm{lab}}}|<1.6$, presented in four ranges of inelasticity $y$ as indicated. The phase-space is further restricted as shown in table \ref{tab:phasespace}.} 
 \label{tab:table_figure_1_d} 
  \end{center} 
   \end{table} 

 \begin{table}[tb] 
 \fontsize{7.5}{7.5}\selectfont
  \begin{center} 
   \begin{tabular}{|cc|c|rrrrrrrrrr|} 
   \multicolumn{13}{c}{\bf \normalsize \boldmath Multiplicity distributions in DIS} \\
   \multicolumn{13}{c}{ \normalsize $5<\qsq<10~\gevsq$ and $-1.2 < \eta_{_{\rm{lab}}} < 0.2$ } \\
   \noalign{\vskip 2mm} 
   \hline 
   \multicolumn{13}{|c|}{ \boldmath $0.0375<y<0.075$} \\
   \hline 
   \multicolumn{2}{|c|}{ } & & & stat. & sys.tot & sys.1 & sys.2 & sys.3 & sys.4 & sys.5 & sys.6 & sys.7 \\ \multicolumn{2}{|c|}{multiplicity range} & $\langle N \rangle$ & $P(N)$ & (\%) & (\%) & (\%) & (\%) & (\%) & (\%) & (\%) & (\%) & (\%) \\
   \hline 
-0.5 & 0.5 & 0      & 0.52 & 0.43 & 13 & 1 & 11 & 5 & 3 & 1 & 1 & 2  \\
0.5 & 1.5 & 1      & 0.29 & 0.84 & 13 & 1 & 11 & 5 & 3 & 1 & 1 & 2  \\
1.5 & 2.5 & 2      & 0.14 & 1.4 & 9.6 & 1 & 9 & 1 & 1 & 1 & 2 & 2  \\
\hline 
 \end{tabular} 
 \caption{Charged particle multiplicity distributions $P(N)$ measured in $ep$ DIS at $\sqrt{s}=319$\gev~with in the ranges of photon virtuality $5<\qsq<10~\gevsq$ and charged particle pseudorapidity range $-1.2<\eta_{_{\rm{lab}}}<0.2$, presented in four ranges of inelasticity $y$ as indicated. The phase-space is further restricted as shown in table \ref{tab:phasespace}.} 
 \label{tab:table_figure_2_a} 
  \end{center} 
   \end{table} 

 \begin{table}[tb] 
 \fontsize{7.5}{7.5}\selectfont
  \begin{center} 
   \begin{tabular}{|cc|c|rrrrrrrrrr|} 
   \multicolumn{13}{c}{\bf \normalsize \boldmath Multiplicity distributions in DIS} \\
   \multicolumn{13}{c}{ \normalsize $5<\qsq<10~\gevsq$ and $-0.5 < \eta_{_{\rm{lab}}} < 0.9$ } \\
   \noalign{\vskip 2mm} 
   \hline 
   \multicolumn{13}{|c|}{ \boldmath $0.0375<y<0.075$ } \\
   \hline 
   \multicolumn{2}{|c|}{ } & & & stat. & sys.tot & sys.1 & sys.2 & sys.3 & sys.4 & sys.5 & sys.6 & sys.7 \\ \multicolumn{2}{|c|}{multiplicity range} & $\langle N \rangle$ & $P(N)$ & (\%) & (\%) & (\%) & (\%) & (\%) & (\%) & (\%) & (\%) & (\%) \\
   \hline 
-0.5 & 0.5 & 0      & 0.18 & 1.3 & 24 & 1 & 23 & 5 & 3 & 1 & 1 & 1  \\
0.5 & 1.5 & 1      & 0.27 & 0.96 & 24 & 1 & 23 & 5 & 3 & 1 & 1 & 1  \\
1.5 & 2.5 & 2      & 0.27 & 0.99 & 18 & 1 & 18 & 1 & 1 & 1 & 2 & 1  \\
\hline 
 \end{tabular} 
 \caption{Charged particle multiplicity distributions $P(N)$ measured in $ep$ DIS at $\sqrt{s}=319$\gev~with in the ranges of photon virtuality $5<\qsq<10~\gevsq$ and charged particle pseudorapidity range $-0.5<\eta_{_{\rm{lab}}}<0.9$, presented in four ranges of inelasticity $y$ as indicated. The phase-space is further restricted as shown in table \ref{tab:phasespace}.} 
 \label{tab:table_figure_2_b} 
  \end{center} 
   \end{table} 

 \begin{table}[tb] 
 \fontsize{7.5}{7.5}\selectfont
  \begin{center} 
   \begin{tabular}{|cc|c|rrrrrrrrrr|} 
   \multicolumn{13}{c}{\bf \normalsize \boldmath Multiplicity distributions in DIS} \\
   \multicolumn{13}{c}{ \normalsize $5<\qsq<10~\gevsq$ and $0.2 < \eta_{_{\rm{lab}}} < 1.6$ } \\
   \noalign{\vskip 2mm} 
   \hline 
   \multicolumn{13}{|c|}{ \boldmath $0.0375<y<0.075$ } \\
   \hline 
   \multicolumn{2}{|c|}{ } & & & stat. & sys.tot & sys.1 & sys.2 & sys.3 & sys.4 & sys.5 & sys.6 & sys.7 \\ \multicolumn{2}{|c|}{multiplicity range} & $\langle N \rangle$ & $P(N)$ & (\%) & (\%) & (\%) & (\%) & (\%) & (\%) & (\%) & (\%) & (\%) \\
   \hline 
-0.5 & 0.5 & 0      & 0.1 & 2.4 & 6.5 & 1 & 2 & 5 & 3 & 1 & 1 & 1  \\
0.5 & 1.5 & 1      & 0.19 & 1.4 & 6.5 & 1 & 2 & 5 & 3 & 1 & 1 & 1  \\
1.5 & 2.5 & 2      & 0.21 & 1.4 & 3.6 & 1 & 2 & 1 & 1 & 1 & 2 & 1  \\
\hline 
 \end{tabular} 
 \caption{Charged particle multiplicity distributions $P(N)$ measured in $ep$ DIS at $\sqrt{s}=319$\gev~with in the ranges of photon virtuality $5<\qsq<10~\gevsq$ and charged particle pseudorapidity range $0.2<\eta_{_{\rm{lab}}}<1.6$, presented in four ranges of inelasticity $y$ as indicated. The phase-space is further restricted as shown in table \ref{tab:phasespace}.} 
 \label{tab:table_figure_2_c} 
  \end{center} 
   \end{table} 

 \begin{table}[tb] 
 \fontsize{7.5}{7.5}\selectfont
  \begin{center} 
   \begin{tabular}{|cc|c|rrrrrrrrrr|} 
   \multicolumn{13}{c}{\bf \normalsize \boldmath Multiplicity distributions in DIS} \\
   \multicolumn{13}{c}{ \normalsize $10<\qsq<20~\gevsq$ and $-1.2 < \eta_{_{\rm{lab}}} < 0.2$ } \\
   \noalign{\vskip 2mm} 
   \hline 
   \multicolumn{13}{|c|}{ \boldmath $0.0375<y<0.075$ } \\
   \hline 
   \multicolumn{2}{|c|}{ } & & & stat. & sys.tot & sys.1 & sys.2 & sys.3 & sys.4 & sys.5 & sys.6 & sys.7 \\ \multicolumn{2}{|c|}{multiplicity range} & $\langle N \rangle$ & $P(N)$ & (\%) & (\%) & (\%) & (\%) & (\%) & (\%) & (\%) & (\%) & (\%) \\
   \hline 
-0.5 & 0.5 & 0      & 0.6 & 0.27 & 13 & 1 & 11 & 5 & 3 & 1 & 1 & 1  \\
0.5 & 1.5 & 1      & 0.26 & 0.66 & 13 & 1 & 11 & 5 & 3 & 1 & 1 & 1  \\
1.5 & 2.5 & 2      & 0.1 & 1.4 & 9.5 & 1 & 9 & 1 & 1 & 1 & 2 & 1  \\
\hline 
 \end{tabular} 
 \caption{Charged particle multiplicity distributions $P(N)$ measured in $ep$ DIS at $\sqrt{s}=319$\gev~with in the ranges of photon virtuality $10<\qsq<20~\gevsq$ and charged particle pseudorapidity range $-1.2<\eta_{_{\rm{lab}}}<0.2$, presented in four ranges of inelasticity $y$ as indicated. The phase-space is further restricted as shown in table \ref{tab:phasespace}.} 
 \label{tab:table_figure_3_a} 
  \end{center} 
   \end{table} 

 \begin{table}[tb] 
 \fontsize{7.5}{7.5}\selectfont
  \begin{center} 
   \begin{tabular}{|cc|c|rrrrrrrrrr|} 
   \multicolumn{13}{c}{\bf \normalsize \boldmath Multiplicity distributions in DIS} \\
   \multicolumn{13}{c}{ \normalsize $10<\qsq<20~\gevsq$ and $-0.5 < \eta_{_{\rm{lab}}} < 0.9$ } \\
   \noalign{\vskip 2mm} 
   \hline 
   \multicolumn{13}{|c|}{ \boldmath $0.0375<y<0.075$ } \\
   \hline 
   \multicolumn{2}{|c|}{ } & & & stat. & sys.tot & sys.1 & sys.2 & sys.3 & sys.4 & sys.5 & sys.6 & sys.7 \\ \multicolumn{2}{|c|}{multiplicity range} & $\langle N \rangle$ & $P(N)$ & (\%) & (\%) & (\%) & (\%) & (\%) & (\%) & (\%) & (\%) & (\%) \\
   \hline 
-0.5 & 0.5 & 0      & 0.21 & 0.8 & 7.9 & 1 & 5 & 5 & 3 & 1 & 1 & 1  \\
0.5 & 1.5 & 1      & 0.26 & 0.73 & 7.9 & 1 & 5 & 5 & 3 & 1 & 1 & 1  \\
1.5 & 2.5 & 2      & 0.25 & 0.79 & 5.8 & 1 & 5 & 1 & 1 & 1 & 2 & 1  \\
\hline 
 \end{tabular} 
 \caption{Charged particle multiplicity distributions $P(N)$ measured in $ep$ DIS at $\sqrt{s}=319$\gev~with in the ranges of photon virtuality $10<\qsq<20~\gevsq$ and charged particle pseudorapidity range $-0.5<\eta_{_{\rm{lab}}}<0.9$, presented in four ranges of inelasticity $y$ as indicated. The phase-space is further restricted as shown in table \ref{tab:phasespace}.} 
 \label{tab:table_figure_3_b} 
  \end{center} 
   \end{table} 

 \begin{table}[tb] 
 \fontsize{7.5}{7.5}\selectfont
  \begin{center} 
   \begin{tabular}{|cc|c|rrrrrrrrrr|} 
   \multicolumn{13}{c}{\bf \normalsize \boldmath Multiplicity distributions in DIS} \\
   \multicolumn{13}{c}{ \normalsize $10<\qsq<20~\gevsq$ and $0.2 < \eta_{_{\rm{lab}}} < 1.6$ } \\
   \noalign{\vskip 2mm} 
   \hline 
   \multicolumn{13}{|c|}{ \boldmath $0.0375<y<0.075$ } \\
   \hline 
   \multicolumn{2}{|c|}{ } & & & stat. & sys.tot & sys.1 & sys.2 & sys.3 & sys.4 & sys.5 & sys.6 & sys.7 \\ \multicolumn{2}{|c|}{multiplicity range} & $\langle N \rangle$ & $P(N)$ & (\%) & (\%) & (\%) & (\%) & (\%) & (\%) & (\%) & (\%) & (\%) \\
   \hline 
-0.5 & 0.5 & 0      & 0.075 & 2 & 6.5 & 1 & 2 & 5 & 3 & 1 & 1 & 1  \\
0.5 & 1.5 & 1      & 0.16 & 1.2 & 6.5 & 1 & 2 & 5 & 3 & 1 & 1 & 1  \\
1.5 & 2.5 & 2      & 0.22 & 1.1 & 3.6 & 1 & 2 & 1 & 1 & 1 & 2 & 1  \\
\hline 
 \end{tabular} 
 \caption{Charged particle multiplicity distributions $P(N)$ measured in $ep$ DIS at $\sqrt{s}=319$\gev~with in the ranges of photon virtuality $10<\qsq<20~\gevsq$ and charged particle pseudorapidity range $0.2<\eta_{_{\rm{lab}}}<1.6$, presented in four ranges of inelasticity $y$ as indicated. The phase-space is further restricted as shown in table \ref{tab:phasespace}.} 
 \label{tab:table_figure_3_c} 
  \end{center} 
   \end{table} 

 \begin{table}[tb] 
 \fontsize{7.5}{7.5}\selectfont
  \begin{center} 
   \begin{tabular}{|cc|c|rrrrrrrrrr|} 
   \multicolumn{13}{c}{\bf \normalsize \boldmath Multiplicity distributions in DIS} \\
   \multicolumn{13}{c}{ \normalsize $20<\qsq<40~\gevsq$ and $-1.2 < \eta_{_{\rm{lab}}} < 0.2$ } \\
   \noalign{\vskip 2mm} 
   \hline 
   \multicolumn{13}{|c|}{ \boldmath $0.0375<y<0.075$ } \\
   \hline 
   \multicolumn{2}{|c|}{ } & & & stat. & sys.tot & sys.1 & sys.2 & sys.3 & sys.4 & sys.5 & sys.6 & sys.7 \\ \multicolumn{2}{|c|}{multiplicity range} & $\langle N \rangle$ & $P(N)$ & (\%) & (\%) & (\%) & (\%) & (\%) & (\%) & (\%) & (\%) & (\%) \\
   \hline 
-0.5 & 0.5 & 0      & 0.76 & 0.24 & 7.9 & 1 & 5 & 5 & 3 & 1 & 1 & 1  \\
0.5 & 1.5 & 1      & 0.18 & 1.1 & 7.9 & 1 & 5 & 5 & 3 & 1 & 1 & 1  \\
1.5 & 2.5 & 2      & 0.045 & 2.9 & 5.8 & 1 & 5 & 1 & 1 & 1 & 2 & 1  \\
\hline 
 \end{tabular} 
 \caption{Charged particle multiplicity distributions $P(N)$ measured in $ep$ DIS at $\sqrt{s}=319$\gev~with in the ranges of photon virtuality $20<\qsq<40~\gevsq$ and charged particle pseudorapidity range $-1.2<\eta_{_{\rm{lab}}}<0.2$, presented in four ranges of inelasticity $y$ as indicated. The phase-space is further restricted as shown in table \ref{tab:phasespace}.} 
 \label{tab:table_figure_4_a} 
  \end{center} 
   \end{table} 

 \begin{table}[tb] 
 \fontsize{7.5}{7.5}\selectfont
  \begin{center} 
   \begin{tabular}{|cc|c|rrrrrrrrrr|} 
   \multicolumn{13}{c}{\bf \normalsize \boldmath Multiplicity distributions in DIS} \\
   \multicolumn{13}{c}{ \normalsize $20<\qsq<40~\gevsq$ and $-0.5 < \eta_{_{\rm{lab}}} < 0.9$ } \\
   \noalign{\vskip 2mm} 
   \hline 
   \multicolumn{13}{|c|}{ \boldmath $0.0375<y<0.075$ } \\
   \hline 
   \multicolumn{2}{|c|}{ } & & & stat. & sys.tot & sys.1 & sys.2 & sys.3 & sys.4 & sys.5 & sys.6 & sys.7 \\ \multicolumn{2}{|c|}{multiplicity range} & $\langle N \rangle$ & $P(N)$ & (\%) & (\%) & (\%) & (\%) & (\%) & (\%) & (\%) & (\%) & (\%) \\
   \hline 
-0.5 & 0.5 & 0      & 0.27 & 0.8 & 7.9 & 1 & 5 & 5 & 3 & 1 & 1 & 1  \\
0.5 & 1.5 & 1      & 0.27 & 0.92 & 7.9 & 1 & 5 & 5 & 3 & 1 & 1 & 1  \\
1.5 & 2.5 & 2      & 0.23 & 1.1 & 5.8 & 1 & 5 & 1 & 1 & 1 & 2 & 1  \\
\hline 
 \end{tabular} 
 \caption{Charged particle multiplicity distributions $P(N)$ measured in $ep$ DIS at $\sqrt{s}=319$\gev~with in the ranges of photon virtuality $20<\qsq<40~\gevsq$ and charged particle pseudorapidity range $-0.5<\eta_{_{\rm{lab}}}<0.9$, presented in four ranges of inelasticity $y$ as indicated. The phase-space is further restricted as shown in table \ref{tab:phasespace}.} 
 \label{tab:table_figure_4_b} 
  \end{center} 
   \end{table} 

 \begin{table}[tb] 
 \fontsize{7.5}{7.5}\selectfont
  \begin{center} 
   \begin{tabular}{|cc|c|rrrrrrrrrr|} 
   \multicolumn{13}{c}{\bf \normalsize \boldmath Multiplicity distributions in DIS} \\
   \multicolumn{13}{c}{ \normalsize $20<\qsq<40~\gevsq$ and $0.2 < \eta_{_{\rm{lab}}} < 1.6$ } \\
   \noalign{\vskip 2mm} 
   \hline 
   \multicolumn{13}{|c|}{ \boldmath $0.0375<y<0.075$ } \\
   \hline 
   \multicolumn{2}{|c|}{ } & & & stat. & sys.tot & sys.1 & sys.2 & sys.3 & sys.4 & sys.5 & sys.6 & sys.7 \\ \multicolumn{2}{|c|}{multiplicity range} & $\langle N \rangle$ & $P(N)$ & (\%) & (\%) & (\%) & (\%) & (\%) & (\%) & (\%) & (\%) & (\%) \\
   \hline 
-0.5 & 0.5 & 0      & 0.061 & 3.2 & 6.5 & 1 & 2 & 5 & 3 & 1 & 1 & 1  \\
0.5 & 1.5 & 1      & 0.14 & 2 & 6.5 & 1 & 2 & 5 & 3 & 1 & 1 & 1  \\
1.5 & 2.5 & 2      & 0.2 & 1.6 & 3.6 & 1 & 2 & 1 & 1 & 1 & 2 & 1  \\
\hline 
 \end{tabular} 
 \caption{Charged particle multiplicity distributions $P(N)$ measured in $ep$ DIS at $\sqrt{s}=319$\gev~with in the ranges of photon virtuality $20<\qsq<40~\gevsq$ and charged particle pseudorapidity range $0.2<\eta_{_{\rm{lab}}}<1.6$, presented in four ranges of inelasticity $y$ as indicated. The phase-space is further restricted as shown in table \ref{tab:phasespace}.} 
 \label{tab:table_figure_4_c} 
  \end{center} 
   \end{table} 

 \begin{table}[tb] 
 \fontsize{7.5}{7.5}\selectfont
  \begin{center} 
   \begin{tabular}{|cc|c|rrrrrrrrrr|} 
   \multicolumn{13}{c}{\bf \normalsize \boldmath Multiplicity distributions in DIS} \\
   \multicolumn{13}{c}{ \normalsize $40<\qsq<100~\gevsq$ and $-1.2 < \eta_{_{\rm{lab}}} < 0.2$ } \\
   \noalign{\vskip 2mm} 
   \hline 
   \multicolumn{13}{|c|}{ \boldmath $0.0375<y<0.075$ } \\
   \hline 
   \multicolumn{2}{|c|}{ } & & & stat. & sys.tot & sys.1 & sys.2 & sys.3 & sys.4 & sys.5 & sys.6 & sys.7 \\ \multicolumn{2}{|c|}{multiplicity range} & $\langle N \rangle$ & $P(N)$ & (\%) & (\%) & (\%) & (\%) & (\%) & (\%) & (\%) & (\%) & (\%) \\
   \hline 
-0.5 & 0.5 & 0      & 0.91 & 0.2 & 6.2 & 1 & 1 & 5 & 3 & 1 & 1 & 1  \\
0.5 & 1.5 & 1      & 0.08 & 2.3 & 6.2 & 1 & 1 & 5 & 3 & 1 & 1 & 1  \\
1.5 & 2.5 & 2      & 0.01 & 9.7 & 3.2 & 1 & 1 & 1 & 1 & 1 & 2 & 1  \\
\hline 
 \end{tabular} 
 \caption{Charged particle multiplicity distributions $P(N)$ measured in $ep$ DIS at $\sqrt{s}=319$\gev~with in the ranges of photon virtuality $40<\qsq<100~\gevsq$ and charged particle pseudorapidity range $-1.2<\eta_{_{\rm{lab}}}<0.2$, presented in four ranges of inelasticity $y$ as indicated. The phase-space is further restricted as shown in table \ref{tab:phasespace}.} 
 \label{tab:table_figure_5_a} 
  \end{center} 
   \end{table} 

 \begin{table}[tb] 
 \fontsize{7.5}{7.5}\selectfont
  \begin{center} 
   \begin{tabular}{|cc|c|rrrrrrrrrr|} 
   \multicolumn{13}{c}{\bf \normalsize \boldmath Multiplicity distributions in DIS} \\
   \multicolumn{13}{c}{ \normalsize $40<\qsq<100~\gevsq$ and $-0.5 < \eta_{_{\rm{lab}}} < 0.9$ } \\
   \noalign{\vskip 2mm} 
   \hline 
   \multicolumn{13}{|c|}{ \boldmath $0.0375<y<0.075$ } \\
   \hline 
   \multicolumn{2}{|c|}{ } & & & stat. & sys.tot & sys.1 & sys.2 & sys.3 & sys.4 & sys.5 & sys.6 & sys.7 \\ \multicolumn{2}{|c|}{multiplicity range} & $\langle N \rangle$ & $P(N)$ & (\%) & (\%) & (\%) & (\%) & (\%) & (\%) & (\%) & (\%) & (\%) \\
   \hline 
-0.5 & 0.5 & 0      & 0.48 & 0.54 & 6.2 & 1 & 1 & 5 & 3 & 1 & 1 & 1  \\
0.5 & 1.5 & 1      & 0.26 & 1.1 & 6.2 & 1 & 1 & 5 & 3 & 1 & 1 & 1  \\
1.5 & 2.5 & 2      & 0.14 & 1.8 & 3.2 & 1 & 1 & 1 & 1 & 1 & 2 & 1  \\
\hline 
 \end{tabular} 
 \caption{Charged particle multiplicity distributions $P(N)$ measured in $ep$ DIS at $\sqrt{s}=319$\gev~with in the ranges of photon virtuality $40<\qsq<100~\gevsq$ and charged particle pseudorapidity range $-0.5<\eta_{_{\rm{lab}}}<0.9$, presented in four ranges of inelasticity $y$ as indicated. The phase-space is further restricted as shown in table \ref{tab:phasespace}.} 
 \label{tab:table_figure_5_b} 
  \end{center} 
   \end{table} 

 \begin{table}[tb] 
 \fontsize{7.5}{7.5}\selectfont
  \begin{center} 
   \begin{tabular}{|cc|c|rrrrrrrrrr|} 
   \multicolumn{13}{c}{\bf \normalsize \boldmath Multiplicity distributions in DIS} \\
   \multicolumn{13}{c}{ \normalsize $40<\qsq<100~\gevsq$ and $0.2 < \eta_{_{\rm{lab}}} < 1.6$ } \\
   \noalign{\vskip 2mm} 
   \hline 
   \multicolumn{13}{|c|}{ \boldmath $0.0375<y<0.075$ } \\
   \hline 
   \multicolumn{2}{|c|}{ } & & & stat. & sys.tot & sys.1 & sys.2 & sys.3 & sys.4 & sys.5 & sys.6 & sys.7 \\ \multicolumn{2}{|c|}{multiplicity range} & $\langle N \rangle$ & $P(N)$ & (\%) & (\%) & (\%) & (\%) & (\%) & (\%) & (\%) & (\%) & (\%) \\
   \hline 
-0.5 & 0.5 & 0      & 0.086 & 2.9 & 6.2 & 1 & 1 & 5 & 3 & 1 & 1 & 1  \\
0.5 & 1.5 & 1      & 0.14 & 2.7 & 6.2 & 1 & 1 & 5 & 3 & 1 & 1 & 1  \\
1.5 & 2.5 & 2      & 0.2 & 2.2 & 3.2 & 1 & 1 & 1 & 1 & 1 & 2 & 1  \\
\hline 
 \end{tabular} 
 \caption{Charged particle multiplicity distributions $P(N)$ measured in $ep$ DIS at $\sqrt{s}=319$\gev~with in the ranges of photon virtuality $40<\qsq<100~\gevsq$ and charged particle pseudorapidity range $0.2<\eta_{_{\rm{lab}}}<1.6$, presented in four ranges of inelasticity $y$ as indicated. The phase-space is further restricted as shown in table \ref{tab:phasespace}.} 
 \label{tab:table_figure_5_c} 
  \end{center} 
   \end{table} 

 \begin{table}[tb] 
 \fontsize{7.5}{7.5}\selectfont
  \begin{center} 
   \begin{tabular}{|cc|c|rrrrrrrrrr|} 
   \multicolumn{13}{c}{\bf \normalsize \boldmath Multiplicity distributions in DIS} \\
   \multicolumn{13}{c}{ \normalsize $5<\qsq<10~\gevsq$ and $0<\rap<4$ } \\
   \noalign{\vskip 2mm} 
   \hline 
   \multicolumn{13}{|c|}{ \boldmath $0.0375<y<0.075$ } \\
   \hline 
   \multicolumn{2}{|c|}{ } & & & stat. & sys.tot & sys.1 & sys.2 & sys.3 & sys.4 & sys.5 & sys.6 & sys.7 \\ \multicolumn{2}{|c|}{multiplicity range} & $\langle N \rangle$ & $P(N)$ & (\%) & (\%) & (\%) & (\%) & (\%) & (\%) & (\%) & (\%) & (\%) \\
   \hline 
-0.5 & 0.5 & 0      & 0.072 & 3.5 & 6.5 & 1 & 1 & 5 & 3 & 1 & 1 & 2  \\
0.5 & 1.5 & 1      & 0.15 & 2.2 & 6.5 & 1 & 1 & 5 & 3 & 1 & 1 & 2  \\
1.5 & 2.5 & 2      & 0.21 & 1.9 & 3.6 & 1 & 2 & 1 & 1 & 1 & 2 & 1  \\
\hline 
 \end{tabular} 
 \caption{Charged particle multiplicity distributions $P(N)$ measured in $ep$ DIS at $\sqrt{s}=319$\gev~with in the ranges of photon virtuality $5<\qsq<10~\gevsq$ and charged particle pseudorapidity range $0<\rap<4$, presented in four ranges of inelasticity $y$ as indicated. The phase-space is further restricted as shown in table \ref{tab:phasespace}.} 
 \label{tab:table_figure_6_a} 
  \end{center} 
   \end{table} 

 \begin{table}[tb] 
 \fontsize{7.5}{7.5}\selectfont
  \begin{center} 
   \begin{tabular}{|cc|c|rrrrrrrrrr|} 
   \multicolumn{13}{c}{\bf \normalsize \boldmath Multiplicity distributions in DIS} \\
   \multicolumn{13}{c}{ \normalsize $10<\qsq<20~\gevsq$ and $0<\rap<4$ } \\
   \noalign{\vskip 2mm} 
   \hline 
   \multicolumn{13}{|c|}{ \boldmath $0.0375<y<0.075$ } \\
   \hline 
   \multicolumn{2}{|c|}{ } & & & stat. & sys.tot & sys.1 & sys.2 & sys.3 & sys.4 & sys.5 & sys.6 & sys.7 \\ \multicolumn{2}{|c|}{multiplicity range} & $\langle N \rangle$ & $P(N)$ & (\%) & (\%) & (\%) & (\%) & (\%) & (\%) & (\%) & (\%) & (\%) \\
   \hline 
-0.5 & 0.5 & 0      & 0.068 & 2.6 & 7.1 & 1 & 2 & 5 & 3 & 1 & 1 & 3  \\
0.5 & 1.5 & 1      & 0.14 & 1.8 & 7.1 & 1 & 2 & 5 & 3 & 1 & 1 & 3  \\
1.5 & 2.5 & 2      & 0.21 & 1.5 & 3.6 & 1 & 2 & 1 & 1 & 1 & 2 & 1  \\
\hline 
 \end{tabular} 
 \caption{Charged particle multiplicity distributions $P(N)$ measured in $ep$ DIS at $\sqrt{s}=319$\gev~with in the ranges of photon virtuality $10<\qsq<20~\gevsq$ and charged particle pseudorapidity range $0<\rap<4$, presented in four ranges of inelasticity $y$ as indicated. The phase-space is further restricted as shown in table \ref{tab:phasespace}.} 
 \label{tab:table_figure_6_b} 
  \end{center} 
   \end{table} 
 \begin{table}[tb] 
 \fontsize{7.5}{7.5}\selectfont
  \begin{center} 
   \begin{tabular}{|cc|c|rrrrrrrrrr|} 
   \multicolumn{13}{c}{\bf \normalsize \boldmath Multiplicity distributions in DIS} \\
   \multicolumn{13}{c}{ \normalsize $20<\qsq<40~\gevsq$ and $0<\rap<4$ } \\
   \noalign{\vskip 2mm} 
   \hline 
   \multicolumn{13}{|c|}{ \boldmath $0.0375<y<0.075$ } \\
   \hline 
   \multicolumn{2}{|c|}{ } & & & stat. & sys.tot & sys.1 & sys.2 & sys.3 & sys.4 & sys.5 & sys.6 & sys.7 \\ \multicolumn{2}{|c|}{multiplicity range} & $\langle N \rangle$ & $P(N)$ & (\%) & (\%) & (\%) & (\%) & (\%) & (\%) & (\%) & (\%) & (\%) \\
   \hline 
-0.5 & 0.5 & 0      & 0.073 & 3.2 & 9.3 & 1 & 5 & 5 & 3 & 1 & 1 & 5  \\
0.5 & 1.5 & 1      & 0.14 & 2.7 & 9.3 & 1 & 5 & 5 & 3 & 1 & 1 & 5  \\
1.5 & 2.5 & 2      & 0.21 & 2.1 & 5.8 & 1 & 5 & 1 & 1 & 1 & 2 & 1  \\
\hline 
 \end{tabular} 
 \caption{Charged particle multiplicity distributions $P(N)$ measured in $ep$ DIS at $\sqrt{s}=319$\gev~with in the ranges of photon virtuality $20<\qsq<40~\gevsq$ and charged particle pseudorapidity range $0<\rap<4$, presented in four ranges of inelasticity $y$ as indicated. The phase-space is further restricted as shown in table \ref{tab:phasespace}.} 
 \label{tab:table_figure_6_c} 
  \end{center} 
   \end{table} 
 \begin{table}[tb] 
 \fontsize{7.5}{7.5}\selectfont
  \begin{center} 
   \begin{tabular}{|cc|c|rrrrrrrrrr|} 
   \multicolumn{13}{c}{\bf \normalsize \boldmath Multiplicity distributions in DIS} \\
   \multicolumn{13}{c}{ \normalsize $40<\qsq<100~\gevsq$ and $0<\rap<4$ } \\
   \noalign{\vskip 2mm} 
   \hline 
   \multicolumn{13}{|c|}{ \boldmath $0.0375<y<0.075$ } \\
   \hline 
   \multicolumn{2}{|c|}{ } & & & stat. & sys.tot & sys.1 & sys.2 & sys.3 & sys.4 & sys.5 & sys.6 & sys.7 \\ \multicolumn{2}{|c|}{multiplicity range} & $\langle N \rangle$ & $P(N)$ & (\%) & (\%) & (\%) & (\%) & (\%) & (\%) & (\%) & (\%) & (\%) \\
   \hline 
-0.5 & 0.5 & 0      & 0.11 & 2.9 & 9.3 & 5 & 1 & 5 & 3 & 1 & 1 & 5  \\
0.5 & 1.5 & 1      & 0.15 & 3.6 & 9.3 & 5 & 1 & 5 & 3 & 1 & 1 & 5  \\
1.5 & 2.5 & 2      & 0.22 & 3 & 3.2 & 1 & 1 & 1 & 1 & 1 & 2 & 1  \\
\hline 
 \end{tabular} 
 \caption{Charged particle multiplicity distributions $P(N)$ measured in $ep$ DIS at $\sqrt{s}=319$\gev~with in the ranges of photon virtuality $40<\qsq<100~\gevsq$ and charged particle pseudorapidity range $0<\rap<4$, presented in four ranges of inelasticity $y$ as indicated. The phase-space is further restricted as shown in table \ref{tab:phasespace}.} 
 \label{tab:table_figure_6_d} 
  \end{center} 
   \end{table} 

 \begin{table}[tb] 
 \fontsize{7.5}{7.5}\selectfont
  \begin{center} 
   \begin{tabular}{|c|rrrrrrrrrr|} 
   \multicolumn{11}{c}{\bf \normalsize \boldmath KNO function of multiplicity distributions in DIS} \\
   \multicolumn{11}{c}{ \normalsize $5<\qsq<10~\gevsq$ and $0<\rap<4$ } \\
   \noalign{\vskip 2mm} 
   \hline 
   \multicolumn{11}{|c|}{ \boldmath $0.0375<y<0.075$ } \\
   \hline 
    & & stat. & sys.tot & sys.1 & sys.2 & sys.3 & sys.4 & sys.5 & sys.6 & sys.7  \\ $z$ & $\Psi(z)$ & (\%) & (\%) & (\%) & (\%) & (\%) & (\%) & (\%) & (\%) & (\%)  \\
   \hline 
 0      & 0.22 & 3.5 & 6.5 & 1 & 1 & 5 & 3 & 1 & 1 & 2  \\
0.33      & 0.45 & 2.2 & 6.5 & 1 & 1 & 5 & 3 & 1 & 1 & 2  \\
0.67      & 0.62 & 1.9 & 3.6 & 1 & 2 & 1 & 1 & 1 & 2 & 1  \\
\hline 
 \end{tabular} 
 \caption{KNO function $\Psi(z)$ of charged particle multiplicity distributions measured in $ep$ DIS at $\sqrt{s}=319$\gev~with in the ranges of photon virtuality $5<\qsq<10~\gevsq$ and charged particle pseudorapidity range $0<\rap<4$, presented in four ranges of inelasticity $y$ as indicated. The phase-space is further restricted as shown in table \ref{tab:phasespace}.} 
 \label{tab:table_figure_7_a} 
  \end{center} 
   \end{table} 

 \begin{table}[tb] 
 \fontsize{7.5}{7.5}\selectfont
  \begin{center} 
   \begin{tabular}{|c|rrrrrrrrrr|} 
   \multicolumn{11}{c}{\bf \normalsize \boldmath KNO function of multiplicity distributions in DIS} \\
   \multicolumn{11}{c}{ \normalsize $10<\qsq<20~\gevsq$ and $0<\rap<4$ } \\
   \noalign{\vskip 2mm} 
   \hline 
   \multicolumn{11}{|c|}{ \boldmath $0.0375<y<0.075$ } \\
   \hline 
    & & stat. & sys.tot & sys.1 & sys.2 & sys.3 & sys.4 & sys.5 & sys.6 & sys.7  \\ $z$ & $\Psi(z)$ & (\%) & (\%) & (\%) & (\%) & (\%) & (\%) & (\%) & (\%) & (\%)  \\
   \hline 
 0      & 0.21 & 2.6 & 7.1 & 1 & 2 & 5 & 3 & 1 & 1 & 3  \\
0.33      & 0.42 & 1.8 & 7.1 & 1 & 2 & 5 & 3 & 1 & 1 & 3  \\
0.66      & 0.64 & 1.5 & 7.1 & 1 & 2 & 5 & 3 & 1 & 1 & 3  \\
\hline 
 \end{tabular} 
 \caption{KNO function $\Psi(z)$ of charged particle multiplicity distributions measured in $ep$ DIS at $\sqrt{s}=319$\gev~with in the ranges of photon virtuality $10<\qsq<20~\gevsq$ and charged particle pseudorapidity range $0<\rap<4$, presented in four ranges of inelasticity $y$ as indicated. The phase-space is further restricted as shown in table \ref{tab:phasespace}.} 
 \label{tab:table_figure_7_b} 
  \end{center} 
   \end{table} 

 \begin{table}[tb] 
 \fontsize{7.5}{7.5}\selectfont
  \begin{center} 
   \begin{tabular}{|c|rrrrrrrrrr|} 
   \multicolumn{11}{c}{\bf \normalsize \boldmath KNO function of multiplicity distributions in DIS} \\
   \multicolumn{11}{c}{ \normalsize $20<\qsq<40~\gevsq$ and $0<\rap<4$ } \\
   \noalign{\vskip 2mm} 
   \hline 
   \multicolumn{11}{|c|}{ \boldmath $0.0375<y<0.075$ } \\
   \hline 
    & & stat. & sys.tot & sys.1 & sys.2 & sys.3 & sys.4 & sys.5 & sys.6 & sys.7  \\ $z$ & $\Psi(z)$ & (\%) & (\%) & (\%) & (\%) & (\%) & (\%) & (\%) & (\%) & (\%)  \\
   \hline 
 0      & 0.22 & 3.2 & 9.3 & 1 & 5 & 5 & 3 & 1 & 1 & 5  \\
0.33      & 0.42 & 2.7 & 9.3 & 1 & 5 & 5 & 3 & 1 & 1 & 5  \\
0.65      & 0.65 & 2.1 & 5.8 & 1 & 5 & 1 & 1 & 1 & 2 & 1  \\
\hline 
 \end{tabular} 
 \caption{KNO function $\Psi(z)$ of charged particle multiplicity distributions measured in $ep$ DIS at $\sqrt{s}=319$\gev~with in the ranges of photon virtuality $20<\qsq<40~\gevsq$ and charged particle pseudorapidity range $0<\rap<4$, presented in four ranges of inelasticity $y$ as indicated. The phase-space is further restricted as shown in table \ref{tab:phasespace}.} 
 \label{tab:table_figure_7_c} 
  \end{center} 
   \end{table} 

 \begin{table}[tb] 
 \fontsize{7.5}{7.5}\selectfont
  \begin{center} 
   \begin{tabular}{|c|rrrrrrrrrr|} 
   \multicolumn{11}{c}{\bf \normalsize \boldmath KNO function of multiplicity distributions in DIS} \\
   \multicolumn{11}{c}{ \normalsize $40<\qsq<100~\gevsq$ and $0<\rap<4$ } \\
   \noalign{\vskip 2mm} 
   \hline 
   \multicolumn{11}{|c|}{ \boldmath $0.0375<y<0.075$ } \\
   \hline 
    & & stat. & sys.tot & sys.1 & sys.2 & sys.3 & sys.4 & sys.5 & sys.6 & sys.7  \\ $z$ & $\Psi(z)$ & (\%) & (\%) & (\%) & (\%) & (\%) & (\%) & (\%) & (\%) & (\%)  \\
   \hline 
 0      & 0.32 & 2.9 & 9.3 & 5 & 1 & 5 & 3 & 1 & 1 & 5  \\
0.36      & 0.42 & 3.6 & 9.3 & 5 & 1 & 5 & 3 & 1 & 1 & 5  \\
0.72      & 0.61 & 3 & 3.2 & 1 & 1 & 1 & 1 & 1 & 2 & 1  \\
\hline 
 \end{tabular} 
 \caption{KNO function $\Psi(z)$ of charged particle multiplicity distributions measured in $ep$ DIS at $\sqrt{s}=319$\gev~with in the ranges of photon virtuality $40<\qsq<100~\gevsq$ and charged particle pseudorapidity range $0<\rap<4$, presented in four ranges of inelasticity $y$ as indicated. The phase-space is further restricted as shown in table \ref{tab:phasespace}.} 
 \label{tab:table_figure_7_d} 
  \end{center} 
   \end{table}

\end{document}